\documentclass[journal=jctc,manuscript=article]{achemso}

\usepackage[version=3]{mhchem} % Formula subscripts using \ce{}
\usepackage{xcolor}
\usepackage{hyperref}
\author{Isaac Smith}
\affiliation{Department of Materials Science and Engineering, Northwestern University, Evanston, IL 60208, USA}
\author{Nicholas Pogharian}
\affiliation{Department of Materials Science and Engineering, Northwestern University, Evanston, IL 60208, USA}
\author{Francisco J. Solis}
\affiliation{School of Mathematical and Natural Sciences, Arizona State University, Glendale, AZ  85306, USA}
\author{Trung Dac Nguyen}
\affiliation{Department of Chemical and Biological Engineering, Northwestern University, Evanston, IL 60208, USA}
\author{Monica Olvera de la Cruz}
\affiliation{Department of Materials Science and Engineering, Northwestern University, Evanston, IL 60208, USA}
\email{m-olvera@northwestern.edu}

\title[DIELECTRIC GPU Package Paper]
  {Electric field effects on electrolytes near rough dielectric surfaces by GPU-accelerated code}

\abbreviations{CPU, GPU, BEM-ICC*, MD, HSMA, DOF, LAMMPS}
\keywords{Electrostatics, Dielectric Mismatch, Molecular Dynamics, Iontronics}

\begin{document}

\begin{tocentry}
\includegraphics[width=8.35cm]{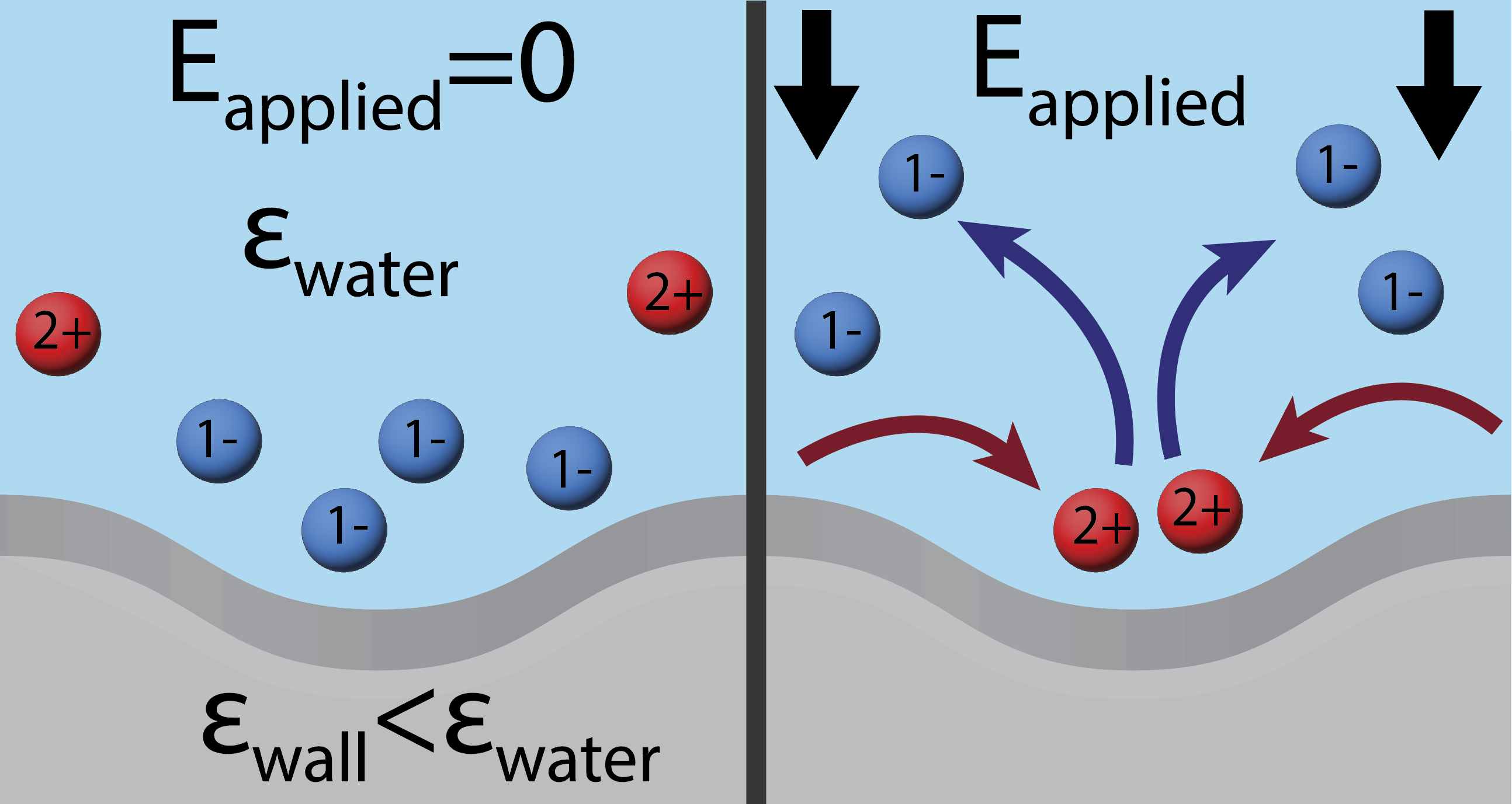}

\end{tocentry}

\begin{abstract}
Dielectric interfaces are ubiquitous in manufactured and natural systems, such as iontronic devices, supercapacitors, and living cells. These, often rough, dielectric surfaces host ionic charge distributions that depend on the surface geometry and the electric fields present. In this work, we study the effect of electric fields on such ionic charge distributions. We demonstrate, by molecular dynamics (MD) and perturbative analytic calculations, that the pattern of alternating regions of ionic charge density created by a sinusoidal interface can be modified and reversed by applying an electric field. We determine the strength of the critical electric field required to cancel the effect of dielectric interface-driven modulation in ion density and develop an analytic expression for that field for small amplitude sinusoidal variations in surface height. We show that ion concentrations near a surface with height given by a sum of Fourier modes can be found by adding the contributions from the concentration modulation due to each mode, allowing the possibility to predict ion distributions near rough surfaces. We updated and validated a LAMMPS package for MD simulation of polarizable surfaces, the DIELECTRIC package, by implementing a new version that achieves a 2.4-24 times speed increase by GPU parallelization on our test system and adds the capability to simulate an applied electric field on simple and complex electrolytes near dielectric interfaces with arbitrary roughness. 
\end{abstract}


\section{Introduction}
Interfaces between materials with different dielectric permittivities are found everywhere in nature and technology. The surface charge that develops at these interfaces in the presence of charges or electric fields guides the behavior of these systems. In life and physical sciences, dielectric interfaces play a key role, for example, in the adsorption \cite{son_image-charge_2021} and transport \cite{dos_santos_modulation_2023} of ions. Dielectric phenomena are important in pores in cell membranes for ion transport \cite{weckel-dahman_ion_2025}, in nanofluidic devices for neuromorphic computing \cite{robin_long-term_2023} and sensing \cite{jia_electricity_2021}, and in porous electrodes for supercapacitor applications \cite{dong_dielectric-electrolyte_2023, eleri_enhanced_2023}.

In many of these systems, such as action potential propagation \cite{subramanian_controlling_2025, marti_action_2018, guo_possible_2024}, supercapacitors \cite{dong_dielectric-electrolyte_2023, eleri_enhanced_2023}, and ionic or iontronic devices, potentially complex dielectric interfaces are coupled to external electric fields. Ionic responses to external electric fields in strongly confined structures and near rough dielectric interfaces are poorly understood.  

Since any continuous periodic function can be represented as a sum of sinusoidal modes, it is natural to begin the study of ions near rough surfaces by considering a sinusoidal dielectric interface in the hope that results from such a system might be generalized through superposition of modes to give insight into the behavior of ions near surfaces with arbitrary shapes. Previous work has analyzed the properties of rough surfaces immersed in ionic solutions without dielectric contrast \cite{bhattacharjee1998dlvo, goldstein1990electric, duplantier1990geometrical, goldstein1991thermodynamics}. A difference in permittivity between the liquid and the solid surface introduces important effects, especially when ions have different valences \cite{levin2009polarizable}. Simulations have shown that the concentration of ions near a sinusoidal dielectric interface is reduced near the troughs and increased near the peaks of the surface, and that this effect is stronger for multivalent ions than for monovalent ones, which leads to a nonzero net ionic charge density near the interface that varies with position. For an asymmetric electrolyte with divalent cations and monovalent anions, above a sinusoidal surface with lower dielectric permittivity than the solvent, the equilibrium ion distribution shows a net negative ionic charge density near the troughs of the sinusoid and a net positive ionic charge density near the peaks of the sinusoid \cite{wu_asymmetric_2018}. Analytic calculations to first-order in the height of a surface deformation have furthermore shown that the interaction of an ion with a sinusoidal dielectric interface is stronger when the ion is in a trough of the surface and weaker when the ion is near a surface peak \cite{fj_pimples_2021}. 

In this work, we investigate how an applied electric field affects the ion distribution near a dielectric surface and how it can be used to manipulate the ionic charge near the surface. We find numerically by molecular dynamics (MD) simulation the critical electric field required to cancel out the modulation in ion density near a sinusoidal dielectric interface, and we show that above this critical field strength the modulation is inverted, going from having lower ionic charge  density in the troughs (and higher at the peaks) to higher ionic charge density in the troughs (and lower at the peaks). We also determine an analytic estimate for the critical field, compare it with results from our MD simulations, and offer an explanation for the discrepancies we found. Finally, we show that the modulation in ion density near a surface with height given by a sum of Fourier modes is the sum of the modulations in ion density due to surfaces with heights given by the individual modes.

Computational study of electric fields near complex dielectric interfaces has been limited by a lack of appropriate methods, and as a result we updated the DIELECTRIC package to increase its speed and add the capability to simulate an applied electric field. Below, we give a brief and non-exhaustive overview of the techniques available for simulating dielectric interfaces, followed by a description of how our code works in the Methods section. (Those who are not interested in the simulation methods can proceed directly to 'Molecular Dynamics' in Methods for simulation details and then to Results and Discussion to read about the developments presented in the preceding paragraph.)

There are several methods for handling dielectric interfaces in MD simulations. Since the induced surface charge at a dielectric interface arises because of the different polarizabilities of the molecules or atoms on either side of the interface, all-atom MD simulations with polarizable atoms can reproduce dielectric interface effects in any geometry. However, these simulations are limited in size, since every part of the system must be filled with polarizable atoms. Even very large polarizable all-atom simulations with millions of atoms only model systems on the length scale of 10-20 nanometers, \cite{adjoua_tinker-hp_2021} and so are not suitable for device simulations.

Implicit solvent MD simulations of ionic solutions explicitly model the ions as discrete objects but treat the solvent and walls or other solid structures as continuous media, each with a dielectric permittivity. Continuum simulations that have included dielectric mismatch have shown strong effects on their physical properties, such as the distributions of and interactions between nanoparticles \cite{barros_dielectric_2014} and droplets \cite{shen_surface_2017}. There are efficient methods for calculating dielectric interactions in MD simulations with flat dielectric interfaces \cite{liang_hsma_2022,yuan_particleparticle_2021}, as well as more computationally demanding methods for any interface that can be meshed. Boundary element methods (BEM) and direct optimization of the energy functional for a system (DOF) \cite{allen_electrostatic_2001,boda_computing_2004, jadhao_simulation_2012} can be used to compute the charge density on a complex surface. It has previously been shown that the BEM with Induced Charge Computation method (BEM-ICC*, a different method from Induced Charge Computation (ICC)\cite{boda_computing_2004}) was faster than DOF or using the generalized minimal residual method with BEM for large systems ($>10^3$ particles) because BEM-ICC* parallelizes well. \cite{nguyen_incorporating_2019}

Current methods for simulating systems with dielectric mismatch that work with externally applied fields are limited to simple cases and/or are computationally expensive. Some of these methods are based on image charges or other methods of obtaining analytic solutions to electrostatics problems for simulating dielectric interfaces with simple geometries or perturbations of simple geometries \cite{robin_modeling_2021, antila_dielectric_2018}. Dos Santos and Levin demonstrated a method based on Green's functions to model ions in a thin slit with dielectric walls \cite{dos_santos_electrolytes_2015}, which can work with electric fields. The method was later extended to work with sinusoidal walls, but only if the amplitude of oscillation is very small, severely limiting the geometries that can be simulated \cite{pogharian_electric_2024}. In some cases, it may be possible to apply an electric field by explicitly adding planar arrangements of charged particles, but since the planes must be infinitely large to create a uniform field, this is impossible in many geometries for many orientations of the field.

In this paper, we extend the BEM-ICC* method that we previously demonstrated \cite{nguyen_incorporating_2019} to work with externally applied electric fields. In the Methods section, we briefly describe how the GPU acceleration of the method is achieved and offer a highlight from the speed comparison, in addition to explaining the boundary condition for the applied electric field. In the Supporting Information, we give more details on the speed increase, derive the boundary condition associated with the method, and validate the method in some static and dynamic cases.

\section{Methods}
\subsection{Enhancements to the DIELECTRIC Package}
In a previous study from our group \cite{nguyen_incorporating_2019}, three methods were implemented to account for the surface polarization effects due to dielectric mismatch between highly immiscible media in coarse-grained MD simulations: (1) the boundary element method (BEM) with the Generalized Minimal Residual solver for Poisson's equation (BEM-GMRES) as proposed by Barros and coworkers \cite{barros_dielectric_2014}, (2) the BEM with image charge computation (BEM-ICC*) by Tyagi and coworkers \cite{tyagi_icc_2010}, and (3) the direct optimization of an energy functional of induced charge density \cite{jadhao_variational_2013}. Our implementations of these methods are released as an optional package called DIELECTRIC in LAMMPS, an open-source massively parallel MD software package \cite{LAMMPS}.

In these field-based methods, the interface is discretized into mesh points, which are modeled as interface particles. The induced charges of the interface particles are computed at the beginning of every MD timestep based on the electric field vectors at each mesh point caused by the charges. All of these calculations were performed on the CPU in the previous study.

\subsubsection{GPU Acceleration}

Here we port the Coulomb and van der Waals interactions between the ions and interface particles to the GPU.  Compared to existing Coulomb models (or \textit{pair\_style} in LAMMPS terminology), our models require a new data structure to store the dielectric constant of the medium where each particle resides.  For interface particles, the difference ($\Delta \epsilon$) and the mean ($\bar{\epsilon}$) of the dielectric constants of the two media are also stored. These per-particle quantities are then copied to the GPU memory for the calculation of Coulomb forces and energies. The electric field at each interface particle due to the real-space contributions from ions and induced charges is also computed on the GPU.  For each particle, multiple GPU threads are scheduled to iterate over the list of its neighboring particles to accumulate the net force and electric field vector at its position. The long-range contribution to the electric field vector and other interactions (such as bonds, angles, and dihedrals) are computed concurrently on the CPU.  In each iteration of the BEM-GMRES and BEM-ICC* methods, the short-ranged parts of the electric field vectors at individual interface particles are transferred from the GPU back to the CPU side to update the induced charges.  The net electric field at each interface particle also includes the effect of the external electric field, if applied.  We expect the performance gain due to GPU acceleration to be high when the interface needs to be discretized into a high-resolution mesh. This is because the iteration over a large number of neighbors per particle for electric field calculation would substantially benefit from the fine-grained parallelism on the GPU.

For a simulation of our test system, 5600 monovalent ions confined between two planar interfaces with 2900 mesh points each, the GPU implementation with 4 MPI ranks was 24 times faster than the CPU implementation with 4 MPI ranks, and was 2.4 times faster than the CPU implementation with 48 MPI ranks, which was the fastest number of MPI ranks for the CPU implementation (we provide more information on the speed comparison in Section S1 of the Supporting Information).

\subsubsection{Applied Electric Fields with BEM-ICC*}

The calculations for computing the induced charge are exactly the same ones that were previously shown for the BEM-ICC* method \cite{nguyen_incorporating_2019}, with one exception: the externally applied electric field is added to the electric field from all the explicit charges in the simulation box in the iterative relaxation for calculating the induced surface charge (this calculation is explicitly presented with the applied electric field in Section S2 of the Supporting Information). During force calculation, with an applied field of $\boldsymbol{E}_\text{applied}$, a particle with charge $q$ feels a force $q\boldsymbol{E}_\text{applied}$ in addition to the forces from the explicitly modeled charges which are computed by the PPPM method. This is exactly analogous to the way the standard \verb|fix efield| command is implemented in LAMMPS. If the applied electric field is uniform in space, then the electrostatic forces on each particle are as if the potential difference across the simulation box were:
\begin{equation}
    \phi(\boldsymbol{x}+\boldsymbol{\hat{x}}_iL_i)-\phi(\boldsymbol{x})=-\boldsymbol{E}_\text{applied}\cdot\boldsymbol{\hat{x}}_iL_i
\end{equation}
where $\boldsymbol{\hat{x}}_i$ and $L_i$ are respectively the unit vector and length of the simulation box along a direction indexed by $i$ and $\phi(\boldsymbol{x})$ is the electric potential at position $\boldsymbol{x}$. This could be described as a "constant potential difference across the simulation box", but while a common interpretation of that phrase for a rectangular box might be that two opposing faces of the box are equipotential surfaces with different potentials, that is not the case here, and this must be kept in mind when considering how the results of simulations performed with this package fit with analytic, experimental, and other numerical results. Validation of the package is presented in Section S3 of the Supporting Information, and a plot of the performance gains from GPU acceleration is presented in Section S1. 

Extra care must be taken if $\boldsymbol{E}_\text{applied}$ has a nonzero component along a direction in which the simulation is periodic. The potential cannot actually be nonperiodic across the simulation box along a periodic direction. So, if a spatially uniform electric field is applied with \verb|fix efield/dielectric| along a periodic direction of the simulation, the electrostatic forces will be calculated as if the above boundary condition is satisfied, but if a charged particle crosses the periodic boundary, its energies in LAMMPS will not match the boundary condition described above, since they will be periodic. In the simulations presented in this work, the simulation box is periodic in all directions. However, the electric field is only applied in the z direction, and all mobile charged particles are confined by walls that prevent them from moving across the z periodic boundaries, thereby avoiding discontinuities in energy.

\subsection{Molecular Dynamics}
MD simulations were run in LAMMPS in reduced units\cite{LAMMPS}. Ions were represented as beads, all with the same radius $\sigma$, which was equal to the Bjerrum length for monovalent ions, and the same mass $m$. All length measurements are given in units of $\sigma$ unless otherwise noted. The energy scale was $\epsilon=e^2/(4\pi\varepsilon_0 \sigma)$, where $e$ is the electron charge and $\varepsilon_0$ is the permittivity of free space. For the dynamic simulations, a timestep of $0.05\tau$ was used, where $\tau=\sqrt{m\sigma^2/\epsilon}$. The hard-core repulsion between particles was modeled by the Weeks-Chandler-Anderson potential\cite{weeks_role_1971}, scaled by a factor of $1/80$ for the relative permittivity of water $\varepsilon_\text{water}=80$:
\begin{equation}
    U_\text{WCA}(r)=\begin{cases}
    \dfrac{\epsilon}{20}\Big[\left(\dfrac{s}{r}\right)^{12}-\left(\dfrac{s}{r}\right)^6\Big]+\epsilon & r\leq2^{1/6}s \\
    0 & r>2^{1/6}s
\end{cases},
\end{equation}
with $s=\sigma$ for ion-ion interactions and $s=0.5\sigma$ for ion-wall interactions. Time evolution in the NVT ensemble was performed by Langevin dynamics with a damping time of $\tau$, and $k_BT=\epsilon/\varepsilon_\text{water} =\epsilon/80$, where $k_B=1.38\times10^{-23}$ J/K is the Boltzmann constant. Except for the sphere test shown in the appendix, all surfaces used a hexagonal mesh. Electrostatic interactions were computed by the Particle-Particle Particle-Mesh (PPPM) method \cite{hockney_computer_1989}. For simulations without an electric field, the spacing between beads in the surface mesh was $0.4\sigma$, the cutoff for real-space electrostatic computations was $5.5\sigma$, and the k-space grid had 16 points in the x and y directions, and 100 points in the z direction. For simulations with an electric field, we found that a finer mesh was required for an accurate field near the surface; we used a bead spacing of $0.1\sigma$, a cutoff of $2.6\sigma$, and a k-space grid with 40 points in the x and y directions and 250 in the z direction. See Section S5 of the Supporting Information for validation of the mesh.

In our simulations, we assumed a constant isotropic dielectric permittivity of water $\varepsilon_\text{water}=80$. In reality, the permittivity of water is strongly anisotropic near (within several nanometers of) interfaces, with the component of the relative permittivity normal to the interface being as low as 2, much lower than the bulk value \cite{fumagalli_anomalously_2018}. We used $\varepsilon_\text{water}=80$ to investigate the effects of dielectric surfaces without introducing the formation of ion pairs or chains, which might occur at low dielectric permittivity, due to stronger interactions between ions.

 To use the GPU-accelerated models, one needs to enable both our new DIELECTRIC package and the LAMMPS GPU package\cite{Brown11,Brown12,Brown13,Trung15,Trung17,Nikolskiy19} in the build. In the present study, we built LAMMPS (2Aug2023) with the GNU GCC 11.2.0 compiler, OpenMPI 4.1.4, and CUDA Toolkit 12.0.1. The GPU package was built with the CUDA backend in mixed-precision mode. Simulations were performed on the Quest cluster at Northwestern University. Each compute node has a 52-core Intel Xeon Platinum 8479N CPU and 2 NVIDIA A30 GPUs.

\section{Results and discussion}

\subsection{Ions Near a Sinusoidal Dielectric Interface in the Presence of an Electric Field}
The ion concentration in an electrolyte near a dielectric interface is different from the bulk concentration, and if the interface is curved, the concentration is different at different points along the surface with different nearby curvature \cite{wu_asymmetric_2018}. The interaction of ions with the surface and the possible presence of external fields both affect the local density. For aqueous electrolytes and low dielectric constant surfaces, the effective self-interaction of ions mediated by the dielectric interface is always repulsive in nature \cite{fj_pimples_2021} and causes a large concentration depletion\cite{wu_asymmetric_2018}. An applied external field can increase or further deplete the ion concentrations.

Both of these effects are also modulated by the shape of the  surface. The combination of surface interaction and external fields can therefore produce a controllable patterning of ionic charge density near the surface. The two effects are not independent, but they can be examined separately to obtain estimates of their relative contributions to the deviation from the bulk ion density. In particular, it is possible to estimate the conditions under which their modulating amplitudes may cancel.

Near the surface, the local mean number density $n$ of a species of valence $Z$ can be considered to arise from the presence of a mean potential $\phi_M$ generated by all other free charges in addition to its self-interaction with the surface $\phi_S$:
\begin{equation}
    n(\mathbf{x})\approx n_B\exp[-(Z\phi_M+Z^2{\phi}_S)/k_BT],
\end{equation}
where $n_B$ is the bulk number density of the species. We have factorized the valence dependence in both terms. In addition to coupling differently with the ion valences, the behavior of the two potentials depends on the surface geometry, electrolyte concentration, and dielectric contrast in different ways. We evaluate the leading contributions to both potentials.

\begin{figure}[H]
    \centering
    \includegraphics[width=\linewidth]{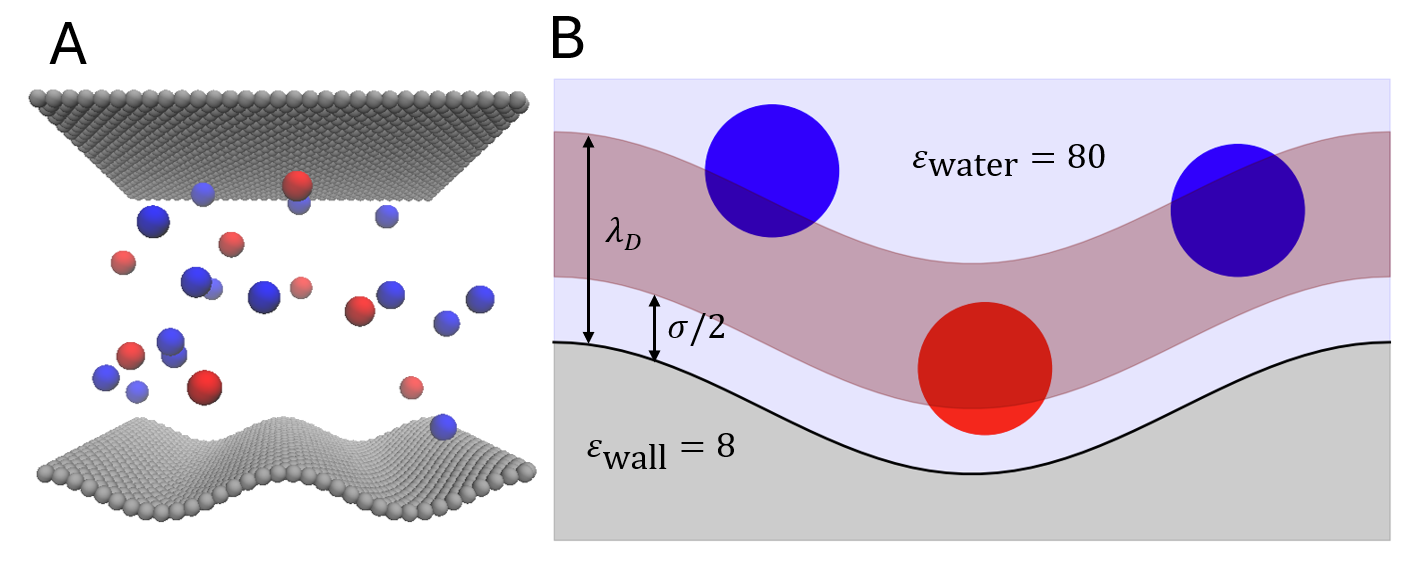}
    \caption{The system being considered here. For the MD results, we used a 2:1 electrolyte with divalent cations. The analytic results are for a generic dilute electrolyte. (A) A frame from an MD simulation of the system. The average distance between the lower and upper walls is $10\sigma$, and there are 54 ions inside the slit. The box shown here has total dimensions $L_x\times L_y\times L_z=6\pi\sigma \times 18.33\sigma\times100\sigma$. The exact dimensions in the $x$ and $y$ dimensions and the number of ions vary with different wavelengths, but the average height of the slit is $10\sigma$ and $L_z=100\sigma$ for all MD results presented in this paper. Exact dimensions for all simulations are given in Section S6 of the Supporting Information. The dielectric permittivity is 80 inside the slit and 8 in the walls outside. (B) A close up of the sinusoidal interface. The shaded region, which is what we consider "near the wall," is the region within 1 Debye length of the surface, excluding the space within an ionic radius of the wall, which ions cannot enter because of steric effects. Both of these figures show a sinusoidal wall height given by $h(x)=0.5\cos(x)$, but we also consider other amplitudes and wavelengths for the sinusoidal surface.}
    \label{fig:system_explanation}
\end{figure}

Here, we examine the ion density near the sinusoidal interface for a 2:1 electrolyte confined between a sinusoidal dielectric interface below and a flat dielectric interface above (Fig \ref{fig:system_explanation}). We first study the system analytically and then compare our findings with the results of our MD simulations.

\subsubsection{Potential due to the Dielectric Self-Interaction}
First we examine an ion's interaction with the dielectric interface due to the surface charge it induces. To start, consider the self-interaction in the presence of a flat surface. A particle of charge $q=Ze$ is in a liquid medium of permittivity $
\varepsilon_B$, while the nearby solid surface has permittivity $
\varepsilon_A$.  We will use the ratio $\gamma=\Delta\varepsilon/\bar{\varepsilon}$,
with $\Delta\varepsilon=\varepsilon_{B}-\varepsilon_{A}$ and $\bar{\varepsilon}=(\varepsilon_{B}+\varepsilon_{A})/2$. The interaction energy of a particle of charge $q$ at a distance $d$ from the surface with the polarization it induces can be described by an image charge $q'=
\gamma q/2$ . The energy of interaction with a flat surface $U_S^{(0)}$ is 

\begin{equation}
    \frac{U_S^{(0)}}{k_BT}=
    \frac{1}{2}\frac{1}{4\pi\varepsilon_0\varepsilon_Bk_BT}\frac{1}{2d}qq'=
    \frac{\gamma}{8}\ell_B\frac{Z^2}{d}.
\end{equation}

Here,
$\ell_{B}=e^{2}/(4\pi\varepsilon_{0}\varepsilon_{B}k_{B}T)$
is the Bjerrum length of the aqueous medium, with $e$ the electron
charge, $\varepsilon_{0}$ the permittivity of free space, and $T$ the temperature.

As reported in previous work in our group, the excess energy $\Delta U_S$ of a single ion near a deformed surface with height $h(x)=A\cos(kx)$ is modulated and has an expression, for short distances, of the form \cite{fj_pimples_2021}

\begin{equation}
\frac{\Delta U_{S}}{k_{B}T}=Z^{2}\ell_{B}A\cos(kx)[\frac{1}{32}\gamma^{2}k^{2}K_0(kz_s)-\frac{\gamma}{8z_s^2}(kz_sK_1(kz_s)-1)] = A\cos(kx)U_S^{(1)}(k, z_s)
\end{equation}
where $z_s$ is the distance from the deformed surface (not from the reference plane at $z=0$). The term containing the modified Bessel functions of the second kind $K_0$ and $K_1$ has a net short-distance expansion that starts with a term of the form $|\ln(kz_s)|$ and is always negative and fast decaying. To evaluate the energy contribution for an ion near the surface, we consider a position an atomic radius $\sigma/2$ above the surface, setting $z_s=\sigma/2$. 

\subsubsection{Potential due to an Applied Field}
We now consider the potential due to an external  field perpendicular to the $z=0$ reference plane. A net ionic charge density is produced inside the liquid medium that screens the external field. The net field decays into the bulk of the liquid, but it is also modified by the surface modulation. This effect is asymmetric, pushing ions toward or away from the surface according to their charge. Estimates of the mean field can be obtained by solving the linear Poisson-Boltzmann equation $\nabla^2\phi-\kappa^2 \phi=0$, where the inverse square screening length is 
$\kappa^{2}=(\sum Z_{i}^{2}n_{i})e^{2}/(\varepsilon_{0}\varepsilon_{B}k_{B}T)$,
with $n_{i}$ the number density of the $i-$th species with valence
$Z_{i}$. In addition, the normal component of the  displacement field $\varepsilon \mathbf{E}$ and the potential $\phi$ must be continuous at the interface. Details of the solution using a perturbation scheme are presented in the appendix. The solution for the flat surface has the form  $\phi=\Phi_0 e^{-\kappa z}$, with $\Phi_0= E_0/(r\kappa)$ and $r=\varepsilon_B/\varepsilon_A$, and the limiting value of the electric field far into the solid region is $\boldsymbol{E}=E_0\hat{\boldsymbol{z}}$. This potential is not singular and can be evaluated at the surface directly as an approximation to the value at one atomic radius distance from it. 
Therefore, the mean field changes the energy of an ion near the flat surface by an amount $U_M$:
\begin{equation}
    \frac{U_M}{k_BT}=\frac{ZeE_0}{k_BTr \kappa}. 
\end{equation}

The solution in the case of a modulated surface gives, in addition, a modulated contribution of the form 
\begin{equation}
    \frac{\Delta U_M}{k_BT}=-\frac{ZAeE_0}{k_BT}\frac{k-\kappa +\sqrt{k^2+\kappa^2}}{k+r \sqrt{k^2+\kappa^2}}\cos(kx).
\end{equation}
For fields directed into the solid and positive particles, this expression is negative at the troughs of the surface. We note that at high levels of screening its value approaches zero. This is consistent with the fact that a highly screening liquid would act as a conductor and create a constant potential at the surface.      

The effects of the self-energy and the presence of an external field have different characteristics and can produce competing behaviors.  In the case of a liquid with higher permittivity than its enclosing solid, the self-energy is positive and independent of the charge sign. It is proportional to the square of the ion's valence and therefore creates  different depletion for different valence ions. The modulation reduces this interaction at the peaks, while enhancing it at the troughs. On the other hand, the external field couples to the charges' signs and is only proportional to the valence. A field pointing into the solid wall reduces the energy of positive charges at the surface. Selecting the direction of the field can enhance or reduce the population of particles of either charge sign. This interaction is also modulated spatially by the surface shape. The magnitude of the excess energy decreases at peaks and increases at troughs. 

In most practical cases, we expect the interaction of a particle with its image to be dominant over other external fields when the particle is near the surface. However, the modulation of this interaction and the modulation of the external field can cancel each other or produce a cancellation of the modulation in the concentration of one of the species. To estimate the appearance of these effects, we can examine the ratio between the magnitudes of the associated energies:
\begin{equation}
    \frac{\Delta U_M}{\Delta U_S}=\frac{128\pi\varepsilon_B\varepsilon_0 (k-\kappa +\sqrt{k^2+\kappa^2})E_0}{Ze (k+r\sqrt{k^2+\kappa^2})[\gamma^2k^2K_0(ka)-4(\gamma/z_s^2)(kz_sK_1(kz_s)-1)]}.
    \label{eq:switching_result}
\end{equation}
Several trends are  readable from this comparison. When the ratio is set equal to 1, we expect the two sources of energy modulation to cancel out, yielding an ion density near the wall that is independent of $x$. We denote the field to achieve this flattening as $E_c$. It can be seen that $E_c$ is independent of the modulation amplitude and that it increases with the ions' valence. The ratio is not in general monotonic with respect to wavelength. The $\Delta U_S$  term in the denominator goes to zero at both short and long wavelengths, but can have a large value in the intermediate range. On the other hand, the numerator $\Delta U_M$ is zero at long wavelengths but tends to a finite limit for short wavelengths. However, the energy ratio is monotonic over the range of wavelengths we consider under the conditions of our MD simulations. The ratio also contains an implicit dependence on the temperature through the screening length $\kappa^{-1}$; as a result, the required field grows with temperature. Finally, the relative permittivity enters the expression in the ratios $s$ and $r$. The critical field $E_c$ increases with the liquid's permittivity.

We performed MD simulations of particles between a flat surface and a sinusoidal surface, with varying wavelengths to find $E_c$ for cations near the surface in a 2:1 electrolyte with divalent cations. To compare the analytic and MD results, we must consider the different boundary conditions used by each method. The electric field in the analytic expressions is the electric field at infinity in a non-periodic system, whereas the electric field set in the MD simulations is the average electric field over the simulation box in a periodic system. The relationship between the field at infinity $E_0$ and the average field $E_\text{avg}$ required to produce the same potential in and near the slit is linear. We denote the ratio between the two fields $\zeta$ such that $\zeta E_0=E_\text{avg}$. $\zeta$ is a function of the heights of the slit and the box as well as the inverse screening length $\kappa$. For this system, $\zeta=0.9004$; the derivation is given in the appendix. The electric field values plotted in Fig \ref{fig:SwitchingFieldvsParameters} correspond to the electric field at infinity, and therefore the fields for the MD simulation data points plotted in Fig \ref{fig:SwitchingFieldvsParameters} are $1/0.9004$ times the average fields set in the MD simulations.

We found qualitative agreement with the trends predicted by the analytic results, though not with the exact value of the field required to cancel out the modulation of the concentration due to the surface geometry (Fig \ref{fig:SwitchingFieldvsParameters}). $E^\text{sim}_c$ are consistently smaller than $E^\text{analytic}_c$, and the discrepancy increases with the wavenumber k of the surface. We attribute this to reduced screening in the MD simulations that is not accounted for in the analytic model. In the MD simulations, an ion in the trough has fewer nearby neighbors in the $+\boldsymbol{\hat{x}}$ and $-\boldsymbol{\hat{x}}$ directions, because the surface has peaks which exclude ions from occupying those regions. As a result, the screening effects are reduced in the troughs and the electric field required to cancel out the curvature effect in the MD simulations is lower than analytically predicted. The excluded volume interaction effect increases with k, as the troughs become narrower, hence the widening gap.

\begin{figure}
    \centering
    \includegraphics[width=0.8\linewidth]{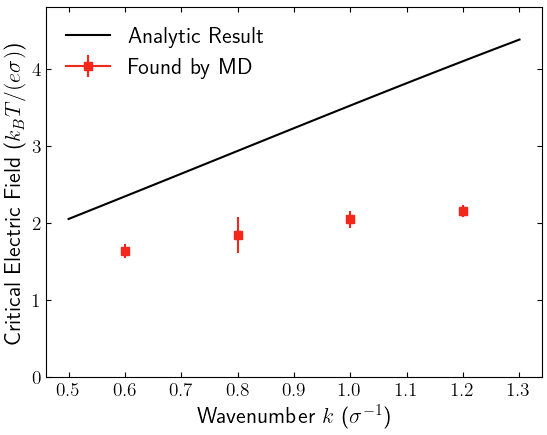}
    \caption{(A) Plot of the electric field $E_c$ required to flatten the cation density near a surface $h(x)=0.5\cos(kx)$ for different values of $k$. The black line is the analytic result from eq \ref{eq:switching_result}, and the red squares are from MD simulations. In the simulations, the relative permittivity is 80 in the solvent and 8 in the walls. For $k=0.6$, the $x$ dimension length of the simulation box was one period of the sinusoid, and for the other wavelengths, it was two periods of the sinusoid, with very similar values in $y$ dimension length. Details on calculation of the exact value of $E_c$ and the error bars\cite{asuero_fitting_2007, taylor_311_1997}, which denote 95\% confidence intervals, are given in Section S7 of the Supporting Information. 20,000 independent samples were used for calculating the critical electric field at each wavenumber.}
    \label{fig:SwitchingFieldvsParameters}
\end{figure}

\begin{figure}
    \centering
    \includegraphics[width=0.8\linewidth]{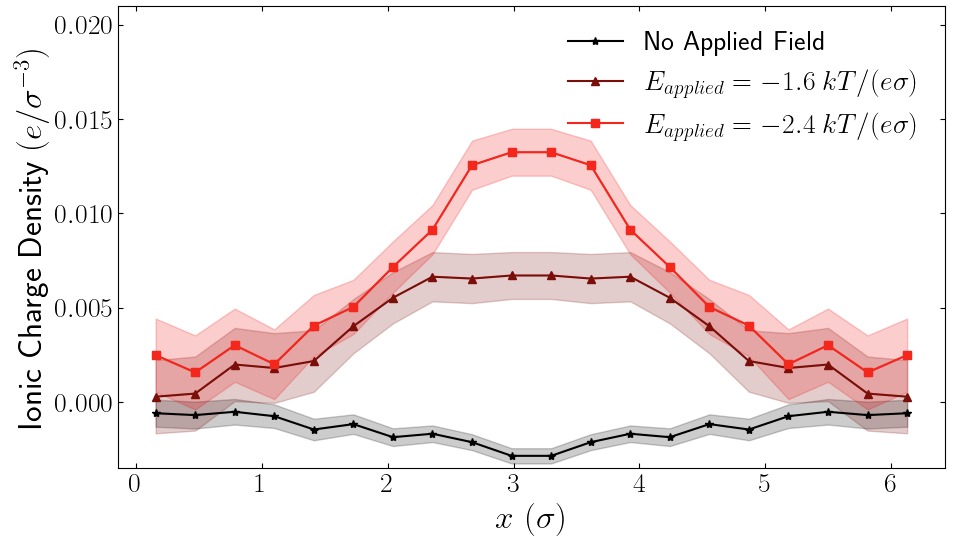}
    \caption{Plot of net ionic charge density from MD simulations showing how an applied electric field can invert and tune the pattern of ionic charge density near a sinusoidal interface. This plot shows net ionic charge density near an interface with height $h(x)=2\cos(x)$. The electrolyte is again a 2:1 salt with 25 mM divalent cation concentration and 50 mM monovalent anion concentration. The electric field values given here are the average value over the box. The shaded regions mark the area within 2 standard errors of the mean charge density at each point. Data is from 8,000 independent samples for the $E_\text{applied}=0$ case and 3,200 independent samples for the cases with applied fields.}
    \label{fig:charge_inversion}
\end{figure}

In addition to determining the field that cancels out the modulating effect of the surface geometry on the concentration of one species of ion, we can also determine the field that cancels out (and then reverses) the modulation of the charge density. In this calculation, we will consider only terms up to first-order in the modulation amplitude $A$ and continue to use the linearized Poisson-Boltzmann equation. Canceling out the modulation of the charge density does not mean that the net charge density will necessarily be zero, only that it will not vary along the length of a sinusoidal period (along the $x$ direction). For a $Z_1$:$Z_2$ electrolyte with $+Z_1$ cations and $-Z_2$ anions (both $Z_1$ and $Z_2$ are positive numbers), the charge density is:
\begin{equation}
    \rho(x)\approx n_B^{(+)}Z_1e\exp((Z_2\phi_M-Z_2^2\phi_S)/k_BT)(\exp(-((Z_1+Z_2)\phi_M+(Z_1^2-Z_2^2)\phi_S)/k_BT)-1)
\end{equation}
The last factor determines where the charge density is positive or negative, and to first-order in the surface height, it is:
\begin{equation}
    -\frac{(Z_1+Z_2)\phi_M+(Z_1^2-Z_2^2)\phi_S}{k_BT}
\end{equation}
If an electrolyte contains ions of different valencies, the charge near the surface will vary along it even without an applied field, and applying an electric field will alter how the charge density varies. This allows for a controllable patterning of charge density in the region near the surface as shown in Fig \ref{fig:charge_inversion}. It should be noted that the charge densities are significantly different than they would be without dielectric contrast, as demonstrated in Section S8 of the Supporting Information.

\subsection{Superposition of Sinusoidal Modes}
Most rough interfaces are not perfect sinusoids, but any bounded, continuous, periodic function can be written in a Fourier series as a (potentially infinite) sum of sinusoidal modes. The height of the deformations in a surface need not be periodic, but we can consider the periodic extension of any finite region. The height also need not be a function (for example, there could be overhangs), but for many surfaces it will be a function, or nearly so.

\begin{figure}
    \centering
    \includegraphics[width=\linewidth]{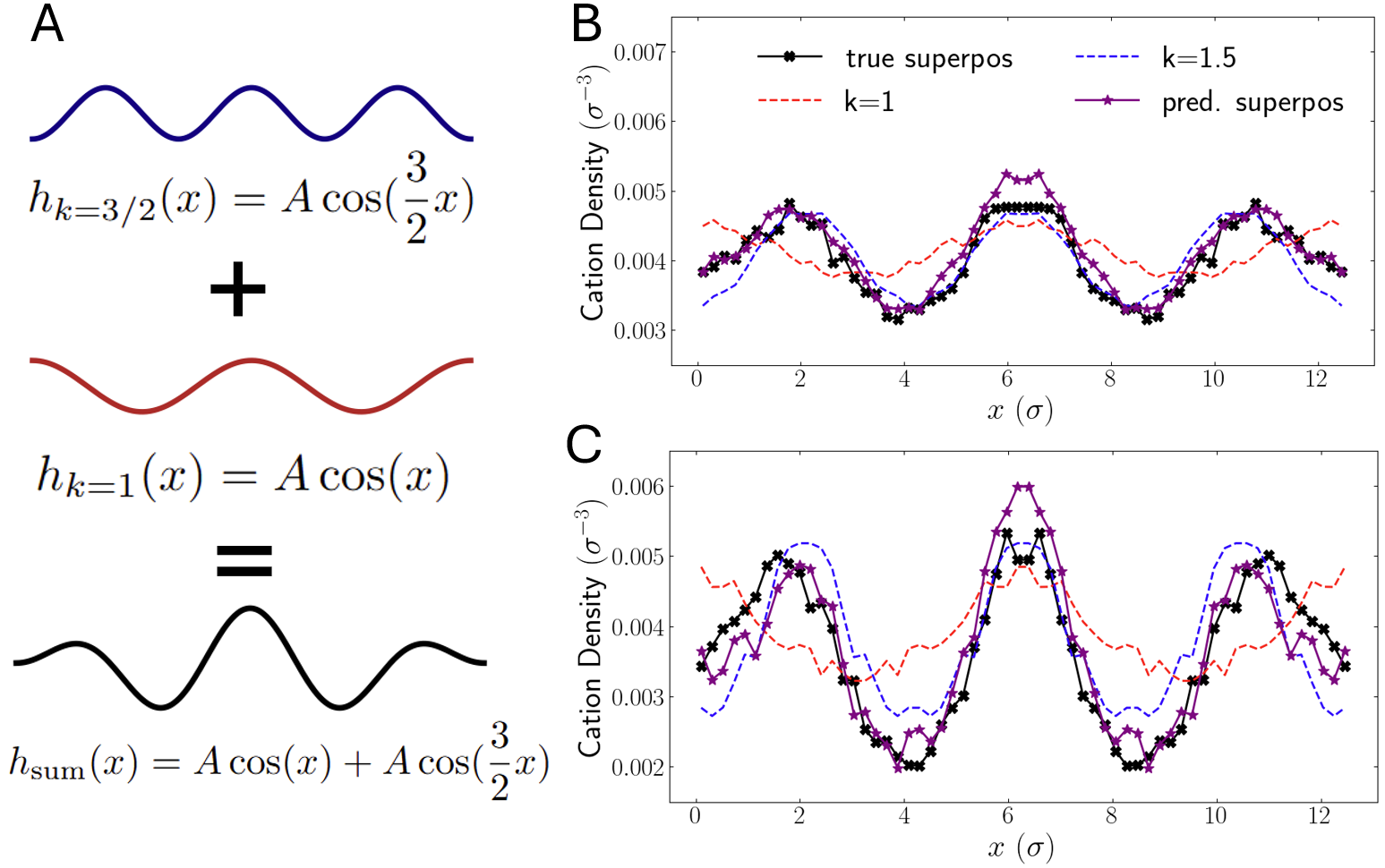}
-    \caption{MD results comparing the ion density near a surface formed by Fourier superposition with the ion densities due to the individual modes that make up that superposition. (A) Schematic showing the two modes ($k=1$, red, and $k=1.5$, blue), and their superposition (black). (B) and (C) Comparison of the modulation in cation density predicted by superposition with the true results from the MD simulation. (B) shows results from the $A=0.25$ case, whereas (C) shows results from the $A=0.5$ case. The density profiles are scaled by bulk concentration to the bulk concentration for the $h_\text{sum}$ system (details given in Section S9 of the Supporting Information). The plots are symmetrical because, for each system, all the data from the simulation was mapped into a single half-period which is then mirrored to show a full period for ease of visualization. In these simulations, the relative permittivity is 80 in the solvent and 2 in the walls. The numbers of independent samples collected in the MD simulations were, for the $A=0.25$ case, 50,000, 30,000, and 40,000 for the $k=1$, $k=1.5$, and $h_\text{sum}$ simulations respectively, and, for the $A=0.5$ case, 30,000, 10,000, and 20,000 for the $k=1$, $k=1.5$, and $h_\text{sum}$ simulations. More samples are required for systems with smaller amplitudes or small $k$ values because these surfaces have lower curvature and so their roughness does not modify the potential near the surface as much.}
    \label{fig:superposition}
\end{figure}

In reality, interfaces can be approximated by a finite number of superposed sinusoidal modes. It is straightforward to extend our analytical approach to understand the response of electrolytes near surface with height given be a superposition of sinusoidal modes (for brevity, we will refer to such a surface as a 'superposed surface'). In this section, we develop a first-order analytic expression for the ion distribution near a superposed surface in terms of the electrolyte distributions that would result from the individual modes that make up that surface. Our expression assumes the potential is linear in surface height, assumes that the ion distribution is linear in potential, and disregards steric interactions, and we compare to MD simulations of the electrolyte distribution near a superposed surface to evaluate the impact of neglecting these nonlinearities. For a surface with height given by $h(x)=\sum_kA_k\cos(kx)$, the dielectric interaction energy is found to be:
\begin{equation}
   U_S = U_S^{(0)}+\Delta U_S=U_0+\sum_kU_S^{(1)}(k,z_s)A_k\cos(kx) 
\end{equation}
For a sum of cosine modes near a wall, if the potentials can also be summed, then, using the expression for $U_S$, and assuming a Boltzmann distribution, the concentration $c(x)$ near the wall is, to first-order in the amplitude:
\begin{equation}
    c(x)= c_\text{bulk}\exp(-\beta U_S)-\sum_k \beta c_\text{bulk}\exp(-\beta U_S^{(0)})U_S^{(1)}(k, z_s)A_k\cos(kx),
\end{equation}
where $c_\text{bulk}$ is the bulk concentration, and $\beta=1/(k_BT)$. For a single cosine mode with wavenumber $k$ and amplitude $A_k$, the concentration would be, to first-order:
\begin{equation}
    c_k(x) = c_\text{bulk}\exp(-\beta U_S)-\beta c_\text{bulk}\exp(-\beta U_S^{(0)}) U^{(1)}_S(k, z_s)A_k\cos(kx), 
\end{equation}
again assuming a Boltzmann distribution. Each single mode alone has a term in the concentration due to the surface height modulation and term that corresponds to the concentration near a flat dielectric interface, which could also be called the $k=0$ mode. For a sum of modes, the term corresponding to a flat dielectric interface should be counted only once overall, not once for each mode, so the near surface concentration for a superposed surface with height given by a sum of $N$ modes with wavenumbers $k_i$ is:
\begin{equation}
    c(x)=-(N-1)c_{k=0}(x)+\sum_i^N c_{k_i}(x),
    \label{eq:superposition_final}
\end{equation}
to first-order in the surface modulation amplitude. This expression is for density profiles from systems with the same bulk concentrations. Due to the mesh used for our dielectric interfaces, the bulk concentrations of our simulations vary across different wavelengths by up to 4.2\%. An explanation of this discrepancy and how we account for it is provided in Section S9 of the Supporting Information. Since ionic charge density is linear in the ion concentrations, the same expression holds for summing ionic charge densities near a surface composed of a sum of sinusoidal modes.
We demonstrate the efficacy of this expression by comparing the concentration of ions near a surface with height $h_\text{sum}(x)=A\cos(x)+A\cos(\frac{3}{2}x)$ with the concentrations near surfaces with heights $h_{k=1}(x)=A\cos(x)$ and $h_{k=3/2}(x) = A\cos(\frac{3}{2}x)$ for the cases of $A=0.25$, and $A=0.5$. Fig \ref{fig:superposition} presents MD results that show that ion density near a surface with height given by a sum of modes is close to the density predicted using the densities due to the individual modes. 
The largest discrepancy between the predicted and true ion densities is at $x=2\pi$ in the $A=0.5$ case, where the total surface height of the $h_\text{sum}$ surface is 1 ion diameter, at which point we expect the first-order description of the potential due to the dielectric interface to start to break down. The superposition of small-amplitude modulations of surface height can create large-amplitude peaks, if several modes interfere constructively, which may bring the superposition out of the regime where the first-order approximation is valid. Additionally, this superposition of the concentrations due to sinusoidal modes with different wavelengths is expected to accurately describe the ion density only when the superposition does not create significant sections of the surface with a radius of curvature smaller than the radius of an ion, since this would create a region where ions cannot enter due to steric effects.

\section{Conclusions}
We investigated the effects of an applied electric field on the distribution of ions near a sinusoidal dielectric interface through MD simulations and analytic calculations. We showed that varying the applied electric field can create a tunable pattern of net ionic charge density across the surface. These effects may be useful in developing nanoscale ionic devices such as memristors and sensors which often involve ions tightly confined near dielectric interfaces \cite{robin_modeling_2021,robin_long-term_2023, dos_santos_modulation_2023}, or patterns of charge along the borders of a channel \cite{curk_discontinuous_2024}, or in altering electrolyte concentrations in the electric double layer to affect catalytic activity \cite{baidoun_recent_2024, gebbie_linking_2023}. If one could determine experimentally the distribution of ions, such as through X-ray standing waves measurements\cite{fenter_electrical_2000}, we could use the distribution of ions to determine where the surface has regions of concave or convex curvature based on where ion concentrations were depleted or enriched compared to the average near the surface, and an applied electric field could enhance these effects for a chosen ion species.

We also showed that the modulation in the density of ions near an interface is nearly a linear combination of the modulations of the ion concentrations due to the Fourier components of that surface. This close-to-linear relationship may help to infer the ion distribution for a rough, non-sinusoidal surface using data from a few sinusoidal surfaces or to design surfaces to achieve a specific ion distribution by combining sinusoidal modes. 

We selected a 2:1 electrolyte because the disproportionately stronger dielectric self-interaction for higher valence ions allows for more complex behavior for an asymmetric electrolyte compared to a symmetric one. In particular, an asymmetric electrolyte is required for a net charge density without an applied electric field. Other asymmetric salts are expected to behave qualitatively similarly to a 2:1 salt. In Section S10 of the Supporting Information, we show some results from the 1:1 salt case. 

To perform our MD simulations, we improved on the DIELECTRIC package by implementing a version of the package's BEM-ICC* method that runs on GPUs, greatly increasing its speed, and by adding the capability of the package to simulate systems with externally applied electric fields. We validated the new version of the package by comparing its output to results from both analytic methods and previously established simulation techniques \cite{pogharian_electric_2024}. This new version of the DIELECTRIC package is ready to use in LAMMPS with dielectric interfaces of arbitrary shape in implicit solvent. In this work, we only simulated fixed interfaces, but the package can also work with mobile dielectric objects or interfaces, though deformable interfaces would require extra consideration to ensure the area for each bead in the surface mesh remains accurate. Although, in principle, a tensorial representation of the water dielectric constant near an interface can be obtained, it is hard to capture the effect of the anisotropy of the permittivity of water for a surface with arbitrary roughness\cite{pogharian_electric_2024}. The present method does not allow for particles to cross dielectric interfaces to simulate the effect of liquid interfaces such as in liquid diodes \cite{jimenez-angeles_functionalities_2025}.

\begin{acknowledgement}

This work was funded by the National Science Foundation (NSF) under Grant No. DMR-2452280. The authors thank Ali Ehlen and Emily Krucker-Velasquez for helpful discussions. The authors thank the Quest High Performance Computing Cluster at Northwestern for computational resources. IS thanks the Hierarchical Materials Cluster Program at Northwestern for funding support through a fellowship.

\end{acknowledgement}


\begin{suppinfo}

Details on the BEM-ICC* method with applied electric field, validation and performance of the code, mesh validation, exact dimensions and concentrations for our systems, notes on uncertainty calculation and superpositions with different concentrations, and brief comparisons to systems with a 1:1 electrolyte and without dielectric contrast.

\end{suppinfo}
\section*{Data and Software Availability}
The data underlying this study are available in the published article, the Supporting Information, and at \href{https://doi.org/10.5281/zenodo.22285686}{https://doi.org/10.5281/zenodo.22285686}. The scripts and code used to generate and analyze the results are also available in the same repository, except for the the new GPU-accelerated version of the DIELECTRIC package, which is available at \href{https://bitbucket.org/NUaztec/lammps-dielectric-gpu/src/main/}{https://bitbucket.org/NUaztec/lammps-dielectric-gpu/src/main/}.

\section{Appendix A: Electric Field Near a Modulated Surface}

In Results and Discussion, we present an expression for the potential near a sinusoidal dielectric interface due to an externally applied electric field. This appendix shows the derivation of that expression and discusses comparing it to our MD results.

\begin{figure}
    \centering
    \includegraphics[width=0.5\linewidth]{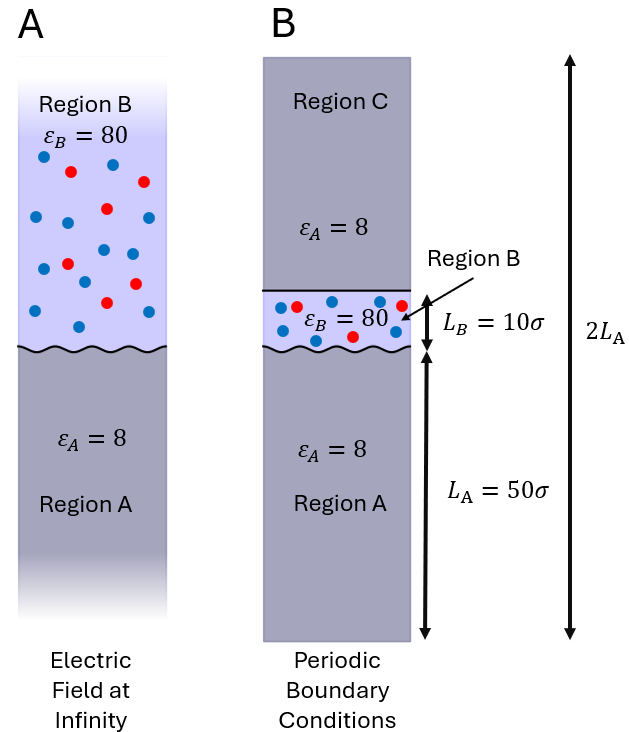}
    \caption{The geometries of the systems for which the solutions are shown in this appendix. (A) shows the geometry for the first solution, where the boundary condition is that the electric field at infinity approaches a specified value. (B) shows the geometry for the second solution, where the value of the electric field is periodic in the $z$ direction, which is the same geometry and boundary condition as in our MD simulations.}
    \label{fig:analytic_systems}
\end{figure}

Modulation in the height of a dielectric interface modifies an otherwise uniform externally applied electric field. We separate this effect from the self-interaction contribution by simply considering the solution to the linear Poisson-Boltzmann equation. Consider a system with permittivity $\varepsilon_A$ in region A, $z<h(x)$, and permittivity $\varepsilon_B$ in region B, $z>h(x)$. $z=h(x)=A\cos(kx)$ is the modulated interface between the two regions. Region A is a solid with no free charges in it, while region B is an electrolyte solution. The system is shown in Fig \ref{fig:analytic_systems}A.

In region A, the Laplace equation $\nabla^2\phi=0$ is satisfied, and the electric field approaches a constant limiting value far into the region. In region B, the linear Poisson-Boltzmann equation $\nabla^2\phi-\kappa^2\phi=0$ is satisfied, where $\kappa$ is the inverse screening length.
At the interface, the potential $\phi$ and the component of the displacement field $\boldsymbol{D}=\varepsilon_0\varepsilon\mathbf{E}$ that is normal to the surface must both be continuous.
We use a perturbation approach and keep only the terms linear in the surface height amplitude $A$. In region A, the potential has the following form, which satisfies the Laplace equation and decays to the prescribed value 
\begin{equation}
    \phi_A=\Phi_0-E_0 z+\alpha \exp(kz)\cos(kx).
\end{equation}
In region B, the potential satisfies the linear Poisson-Boltzmann equation and it has the following form:
\begin{equation}
    \phi_B= \Phi_0 \exp(-\kappa z)+\beta \exp(-\sqrt{k^2+\kappa^2} z)\cos(kx).
\end{equation}
Note that $\alpha$ and $\beta$ will turn out to be proportional to $A$, so the terms including those factors are first-order.
Evaluating the expressions for the potential and field at the deformed interface, we obtain the following relations. Continuity of the potential to first-order at $z=A\cos(kx)$ gives 
\begin{equation}
    \Phi_0-E_0 A\cos(kx)+\alpha\cos(kx)=\Phi_0+\beta\cos(kx)-\kappa \Phi_0 A\cos(kx). 
\end{equation}
Continuity of the displacement field can be checked with just its $z$-component since, in the transversal direction, all first-order contributions vanish.  
\begin{equation}
    \varepsilon_A(E_0-k\alpha\cos(kx))=\varepsilon_B(\kappa\Phi_0-\kappa^2\Phi_0A\cos(kx)+\beta\sqrt{\kappa^2+k^2}\cos(kx)). 
\end{equation}
Solutions to the system are given by the following relations. First, from the non-cosine terms in the continuity of the displacement field, the amplitude of the potential $\Phi_0$ is determined by the external field to be
\begin{equation}
    r\kappa \Phi_0= E_0,
\end{equation}
where $r=\varepsilon_B/\varepsilon_A$
The coefficients $\alpha$ and $\beta$ are:
\begin{eqnarray}
    \alpha&=&E_0A\frac{\kappa+(r-1)\sqrt{k^2+\kappa^2}}{k+r\sqrt{k^2+\kappa^2}},\\
    \beta&=& \frac{E_0A}{r}\frac{r \kappa-(r-1)k}{ k+r\sqrt{k^2+\kappa^2}}.
    \label{eq:first_calc_alpha}
    \label{eq:first_calc_beta}
\end{eqnarray}

Using these expressions, we can obtain an evaluation of the potential near the surface. It can be checked that the following expression has the same first-order coefficients just determined:
\begin{equation}
    \phi=\frac{E_0}{r\kappa}\exp[-\kappa( z-h(x))]-AE_0\frac{ k-\kappa+\sqrt{k^2+\kappa^2}}{k+r\sqrt{k^2+\kappa^2}}\exp[-\sqrt{k^2+\kappa^2}(z-h(x))]\cos(kx).
    \label{eq:first_calc_potential}
\end{equation}

By shifting the evaluation point of the exponentials by the height of the deformed surface $h(x)$, the potential at the surface itself is easily recovered:
\begin{equation}
    \phi_M=\frac{1}{r \kappa }E_0-AE_0\frac{k-\kappa +\sqrt{k^2+\kappa^2}}{k+r \sqrt{k^2+\kappa^2}}\cos(kx)
    \label{eq:surface_phi}
\end{equation}
The first term represents the potential due to a flat dielectric interface, and the second term is the modulation in that potential due to the fluctuating height of the surface.

\subsection{Comparison to MD Simulation Results}

Eq. \ref{eq:surface_phi} is a useful expression for the effect of an electric field, but the system it was derived for does not match the boundary conditions used in the MD simulations. The applied electric field used with the DIELECTRIC package is the average electric field across the simulation box, whereas the applied electric field in the above derivation is the electric field at infinity, for a non-periodic system. We must consider this difference when comparing the analytic result to the results from the MD simulations. The relationship between the macroscopic values of the field, defined by the potential differences at boundaries and the local field values, has been discussed in relation to transient phenomena in electrolyte solutions (the so-called “IR” problem, \cite{bazant2004diffuse})and for the case of oscillatory fields in heterogeneous materials \cite{solis2023electrical}.  This local-to-macroscopic relation is also non-trivial in static cases such as those treated in our numerical work.

It is possible to solve the Poisson and Poisson-Boltzmann equations for the system in the MD simulations with appropriate periodic boundary conditions. We consider a box $-L_A<z<L_A$. Region A, $-L_A<z<h(x)=A\cos(kx)$, has permittivity $\varepsilon_A$. Region B, $h(x)<z<L_B$, has permittivity $\varepsilon_B$, and an electrolyte solution with inverse screening length $\kappa$. And region C, $L_B<z<L_A$, has permittivity $\varepsilon_A$. The electric field is periodic, so $\boldsymbol{E}_A(-L_A)=\boldsymbol{E}_C(+L_A)$, but the potential is not periodic in z, since there is an applied field. The potential is periodic in $x$ with a periodicity of $2\pi/k$, equal to the periodicity of $h(x)$. The system is shown in Fig \ref{fig:analytic_systems}B.

The form of the solution is similar to the one given above, but with a few extra terms. It is:
\begin{align}
    \phi_A = -E_0z + \Phi_1 + \Phi_2e^{-\kappa L_B} + \alpha e^{kz}\cos(kx) +\gamma e^{-k(z+2L_A)}\cos(kx)\\
    \phi_B = \Phi_1 e^{-\kappa z} + \Phi_2 e^{\kappa(z-L_B)}+\beta e^{-\sqrt{k^2+\kappa^2}z}\cos(kx)+\eta e^{\sqrt{k^2+\kappa^2}(z-L_B)}\cos(kx)\\
    \phi_C = -E_0(z-L_B) + \Phi_1 e^{-\kappa L_B} + \Phi_2 + \alpha e^{k(z-2L_A)}\cos(kx) +\gamma e^{-kz}\cos(kx)    
\end{align}

$\phi_A$ solves the Poisson equation in region A, $\phi_B$ solves the linearized Poisson-Boltzmann equation in region B, and $\phi_C$ solves the Poisson equation in region C.
The boundary conditions at the A-B and B-C interfaces are again continuity of the potential and the displacement fields. The boundary condition at the A-C interface, which is due to periodic boundary conditions, requires only continuity of the electric field, and is automatically satisfied by the form of the equations. In the end, these boundary conditions give six equations:
\begin{align}
\varepsilon_A E_0
&= \varepsilon_B\left(\kappa\Phi_1-\kappa\Phi_2e^{-\kappa L_B}\right), \\
\varepsilon_A E_0
&= \varepsilon_B\left(\kappa\Phi_1e^{-\kappa L_B}-\kappa\Phi_2\right), \\
-E_0A+\alpha+\gamma e^{-2kL_A}
&=
-\kappa A\Phi_1+\kappa A\Phi_2e^{-\kappa L_B}
+\beta+\eta e^{-\sqrt{k^2+\kappa^2}L_B}, \\
\varepsilon_A\left(-k\alpha+k\gamma e^{-2kL_A}\right)
&=
\varepsilon_B\Bigl(
-\kappa^2A\Phi_1
-\kappa^2A\Phi_2e^{-\kappa L_B}
\Bigr) \nonumber\\
&\quad
+\varepsilon_B\Bigl(
\beta\sqrt{k^2+\kappa^2}
-\eta\sqrt{k^2+\kappa^2}
e^{-\sqrt{k^2+\kappa^2}L_B}
\Bigr), \\
\alpha e^{k(L_B-2L_A)}+\gamma e^{-2kL_B}
&=
\beta e^{-\sqrt{k^2+\kappa^2}L_B}+\eta, \\
\varepsilon_A\left(-k\alpha e^{k(L_B-2L_A)}
+k\gamma e^{-kL_B}\right)
&=
\varepsilon_B\left(
\beta\sqrt{k^2+\kappa^2}
-\eta\sqrt{k^2+\kappa^2}
e^{-\sqrt{k^2+\kappa^2}L_B}
\right).
\end{align}
The first two equations are easily solved for $\Phi_1$ and $\Phi_2$:

\begin{eqnarray}
    \Phi_1&=&\frac{E_0}{r\kappa(1+e^{-\kappa L_B})}\\
    \Phi_2&=&-\frac{E_0}{r\kappa(1+e^{-\kappa L_B})}
\end{eqnarray}
The solution for the other four coefficients is given by
\begin{equation}
    \begin{pmatrix}
1 &
-1 &
e^{-2kL_A} &
- e^{-\xi L_B}
\\

-k\varepsilon_A &
-\varepsilon_B \xi &
k\varepsilon_A e^{-2kL_A} &
\varepsilon_B \xi e^{-\xi L_B}
\\

e^{k(L_B-2L_A)} &
- e^{-\xi L_B} &
e^{-kL_B} &
-1
\\

k\varepsilon_A e^{k(L_B-2L_A)} &
\varepsilon_B \xi e^{-\xi L_B} &
-k\varepsilon_A e^{-kL_B} &
-\varepsilon_B \xi
\end{pmatrix}
\begin{pmatrix}
\alpha \\
\beta \\
\gamma \\
\eta
\end{pmatrix}
=
\begin{pmatrix}
E_0 A - \kappa A\Phi_1 + \kappa A\Phi_2 e^{-\kappa L_B}\\
-\varepsilon_B\kappa^2 A\Phi_1 - \varepsilon_B\kappa^2 A\Phi_2 e^{-\kappa L_B} \\
0 \\
0
\end{pmatrix}.
\end{equation}
where $\xi \equiv \sqrt{k^2+\kappa^2}$. The explicit expression for the potential is long, and does not provide as much insight into the dependence on $k,\;\kappa$, and $r$ as Eq. \ref{eq:first_calc_potential}, which was computed using the electric field at infinity boundary condition. However, the explicitly computed values for $\Phi_1$ and $\Phi_2$ are enough to determine the average electric field across one periodic box to be:
\begin{equation}
    \boldsymbol{E}_\text{avg}=\frac{E_0\boldsymbol{\hat{z}}}{2L_A}\bigg(\frac{\tanh(\kappa L_B/2)}{r\kappa}+2L_A-L_B\bigg)
\end{equation}
Note that there is no first-order dependence on the modulation amplitude; in fact, this is the same result we would have gotten if we had assumed a flat surface. We define $\zeta$ to be the ratio between the average field and the field at infinity, $E_\text{avg}=\zeta E_0$. For our simulations, $\kappa=0.6$, $L_B=10$, and $L_A=50$, and we find $\zeta=0.9004$. To first-order, the modulation in potential at the surface is:
\begin{equation}
    \Delta \phi_M = \bigg(-\frac{AE_0}{r}\tanh(\kappa L_B/2)+\beta+\eta e^{-\sqrt{k^2+\kappa^2}L_B}\bigg)\cos(kx)
\end{equation}

For both this calculation (with periodic boundary conditions) and the one above (with the electric field specified at infinity), the modulation in the potential at the surface due to the applied field takes the form $CAE_0\cos(kx)$, where $C$ is a constant that depends on $k, \; \kappa$, and $r$. The values of $C$ (Table \ref{tab:C_table}) show that discrepancies between the two different boundary conditions are in the third or fourth significant digit for the parameters we investigated, which is negligible compared to the differences between the MD results and analytic results, which are around 30-40\%. Therefore, we use the potential from the infinite boundary condition case in our comparisons with MD simulations for the sake of an interpretable expression. However, from the derivation with periodic boundary conditions, we see that, in order to test an applied field of $E_0$ at infinity, an average applied field of $E_\text{avg}=0.9004E_0$ should be used in the MD simulation.
\begin{table}
\centering
\begin{tabular}{|l|l|l|l|l|} 
\hline
 & $k=0.6$ & $k=0.8$ & $k=1$ & $k=1.2$\\ 
\hline
$C$ for Electric Field at Infinity & -0.09339 & -0.11111 & -0.12369 & -0.13283 \\ 
\hline
$C$ for Periodic Boundary Conditions & -0.09322 & -0.11089 & -0.12343 & -0.13254\\ 
\hline
\end{tabular}
\caption{Proportionality constants $C$ for the modulation in potential $\Delta \phi_M(x) = CAE_0\cos(kx)$. Parameters are the same as for the MD simulations with electric field: $\kappa=0.6,\;\varepsilon_A=8,\;\varepsilon_B=80$, and $r=10$.}
\label{tab:C_table}
\end{table}
\newpage
\section{\Huge Supporting Information}

\section{Section S1: Code Performance Evaluation}
\begin{figure}
    \centering
    \includegraphics[width=0.8\linewidth]{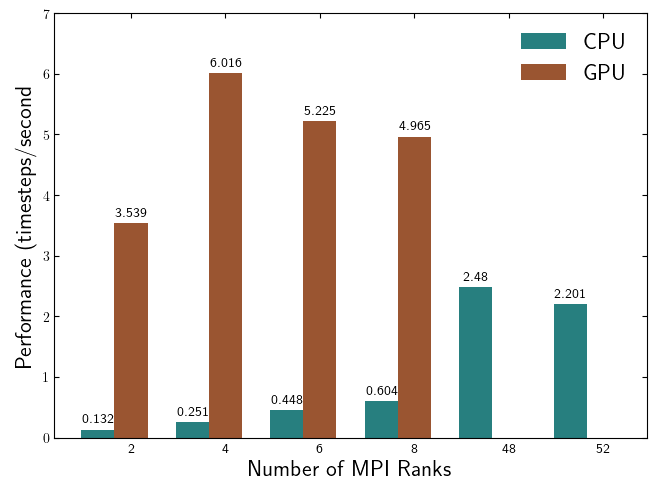}
    \caption{Comparison of the performance of the CPU and GPU versions of the DIELECTRIC package. The CPU implementation was tested at 2, 4, 6, 8, and multiples of 4 up to 52 MPI ranks. The GPU implementation was tested at 2, 4, 6, and 8 MPI ranks each with an equivalent  number of GPUs. We did not test higher numbers of GPUs due to availability of GPUs on the computing cluster. The best performance of the GPU implementation was achieved with 4 MPI ranks, and the GPU implementation at 2, 4, 6, and 8 MPI ranks was substantially faster than the CPU implementation for all the MPI ranks measured (the fastest of which was 48 MPI ranks).}
    \label{fig:performance_comparison}
\end{figure}

To examine the performance of the GPU implementation compared to the CPU-only version, we simulated a system of a 1:1 electrolyte solution confined between two planar interfaces. Each interface was composed of 2900 particles arranged in a triangular lattice. The distance between the two interfaces was $D = 10\sigma$. The 5600 monovalent ions were equilibrated in the canonical ensemble using the Langevin thermostat. The total number of particles in the benchmark system was $N$ = 11400, and the simulation box dimensions were $L_x = L_y = L_z = 50\sigma$. We simulated this system using the GPU implementation and its CPU-based predecessor. The results (Fig \ref{fig:performance_comparison}) show that the GPU implementation is substantially faster, especially for a small number of MPI ranks. Even with only 2 MPI ranks, the GPU implementation was faster than the CPU implementation for all numbers of MPI ranks tested (up to 52). The GPU implementation was fastest at 4 MPI ranks, which was 24 times faster than the CPU implementation with 4 MPI ranks, and 2.42 times faster than the fastest CPU implementation speed recorded.

\section{Section S2. Details of the BEM-ICC* Method with Electric Field}
In this section, we present essentially the same derivation of the BEM-ICC* method that we showed previously\cite{nguyen_incorporating_2019}, but we explicitly explain how the applied electric field fits into the method and show how it leads to the boundary condition we stated in the main text.
Consider two homogeneous media with different dielectric constants $\varepsilon_1$ and $\varepsilon_2$, with a sharp boundary between them, which may also have a fixed surface charge density on it.
Let $\boldsymbol{E}^{(1)}$ be the electric field in the first medium, which we can split into $\boldsymbol{E}^{(1)}=\boldsymbol{E}_q^{(1)}+\boldsymbol{E}_\text{applied}$, where $\boldsymbol{E}_q^{(1)}$ is the field due to all the charges, including fixed and bound charges on the interface, and $\boldsymbol{E}_\text{applied}$ is the externally applied field. We can similarly split the electric field in the second medium: $\boldsymbol{E}^{(2)}=\boldsymbol{E}_q^{(2)}+\boldsymbol{E}_\text{applied}$.
Suppose the dielectric interface is discretized by a mesh, and we consider a point $\boldsymbol{s}$ on that mesh. The discontinuity in the displacement field $\boldsymbol{D}=\varepsilon\varepsilon_0\boldsymbol{E}$, which satisfies $\boldsymbol{\nabla}\cdot\boldsymbol{D}=\rho_f$ at the interface, is due to the fixed surface charge density:
\begin{equation}
    \varepsilon_0[\varepsilon_2\boldsymbol{E}_q^{(2)}(\boldsymbol{s})-\varepsilon_1\boldsymbol{E}_q^{(1)}(\boldsymbol{s})]\cdot \boldsymbol{n}(\boldsymbol{s})+\varepsilon_0\Delta\varepsilon\boldsymbol{E}_\text{applied}=\sigma_f(\boldsymbol{s}),
\end{equation}
where $\boldsymbol{n}(\boldsymbol{s})$ is the unit vector normal to the surface at point $\boldsymbol{s}$ and points into medium 2, and $\Delta\varepsilon = \varepsilon_2-\varepsilon_1$. The discontinuity in the electric field is due to all the charge on the interface:
\begin{equation}
        \varepsilon_0[\boldsymbol{E}^{(2)}(\boldsymbol{s})-\boldsymbol{E}^{(1)}(\boldsymbol{s})]\cdot \boldsymbol{n}(\boldsymbol{s})=\sigma_f(\boldsymbol{s})+\sigma_b(\boldsymbol{s})
\end{equation}
Following Tyagi et al.\cite{tyagi_icc_2010}, the electric fields on either side of the interface are approximated by:
\begin{equation}
    \boldsymbol{E}^{(1)}(\boldsymbol{s})=\boldsymbol{E}(\boldsymbol{s})-\frac{\sigma_b(\boldsymbol{s})+\sigma_f(\boldsymbol{s})}{2\varepsilon_0}\boldsymbol{n}(\boldsymbol{s}), \;\;\;\;
        \boldsymbol{E}^{(2)}(\boldsymbol{s})=\boldsymbol{E}(\boldsymbol{s})+\frac{\sigma_b(\boldsymbol{s})+\sigma_f(\boldsymbol{s})}{2\varepsilon_0}\boldsymbol{n}(\boldsymbol{s})
        \label{eq:e1e2def}
\end{equation}
where
$\boldsymbol{E}(\boldsymbol{s})=\boldsymbol{E}_q(\boldsymbol{s})+\boldsymbol{E}_\text{applied}$ is the electric field at the interface which includes the applied field and the contribution from all explicitly modeled charges. $\boldsymbol{E}^{(1)}$ and $\boldsymbol{E}^{(2)}$ can be substituted into eq (\ref{eq:e1e2def}) to obtain:
\begin{equation}
    \sigma_b(\boldsymbol{s})=\frac{1-\bar{\varepsilon}}{\bar{\varepsilon}}\sigma_f(\boldsymbol{s})-\varepsilon_c\Delta\frac{\varepsilon}{\bar{\varepsilon}}(\boldsymbol{E}_q(\boldsymbol{s})+\boldsymbol{E}_\text{applied})\cdot\boldsymbol{n}(\boldsymbol{s})
\end{equation}
where $\bar{\varepsilon}=(\varepsilon_1+\varepsilon_2)/2$. The surface charge density is then solved for iteratively. Just as we split the total field as $\boldsymbol{E}=\boldsymbol{E}_q+\boldsymbol{E}_\text{applied}$, we can split the potential:
$\phi=\phi_q+\phi_\text{applied}$, where $\phi_q$ is due to the charges in the box, and $\phi_\text{applied}$ is due to the applied field. Since $\boldsymbol{E}_q$ is due to the explicitly modeled charged beads and computed by the PPPM method, the associated potential is periodic over the box length, so $\phi_q(\boldsymbol{x})=\phi_q(\boldsymbol{x}+\boldsymbol{\hat{x}}_iL_i)$, for each direction $i$, which has box dimension $L_i$ and unit vector $\boldsymbol{\hat{x}}_i$. Then the total potential difference over a box length is only due to the difference in $\phi_\text{applied}$. If $\boldsymbol{E}_\text{applied}$ is constant over the entire space of the box, we can derive a boundary condition from the potential difference across the box.
\begin{equation}
    \phi_\text{applied}(\boldsymbol{x}+\boldsymbol{\hat{x}}_iL_i)-\phi_\text{applied}(\boldsymbol{x})=
    -\int_{\boldsymbol{x}} ^{\boldsymbol{x}+\boldsymbol{\hat{x}}_iL_i}\boldsymbol{E}_\text{applied}\cdot\boldsymbol{\hat{x}}_idl = -\boldsymbol{E}_\text{applied}\cdot\boldsymbol{\hat{x}}_iL_i
\end{equation}
So electric fields felt by particles are calculated as if the boundary condition imposed were $    \phi(\boldsymbol{x}+\boldsymbol{\hat{x}}_iL_i)-\phi(\boldsymbol{x})=-\boldsymbol{E}_\text{applied}\cdot\boldsymbol{\hat{x}}_iL_i$. As we mention in the main text, this does not actually mean the potential can be nonperiodic across the simulation box in a periodic direction, and care must be taken if an external field is applied in a periodic direction along which particles can cross the periodic boundary.

The DIELECTRIC package also allows the user to apply an electric field that varies in space, in which case, the boundary conditions would have to be examined on a case-by-case basis. In any event, for a physically sensible simulation, the user-chosen applied field should be one that could be generated by charges outside the simulation box (it must have zero divergence inside the simulation box and any of its periodic images) and should be periodic along any periodic direction of the simulation.

\section{Section S3. Code Validation}

\subsection{Validation for Static Cases}
\begin{figure}
    \centering
    \includegraphics[width=1\linewidth]{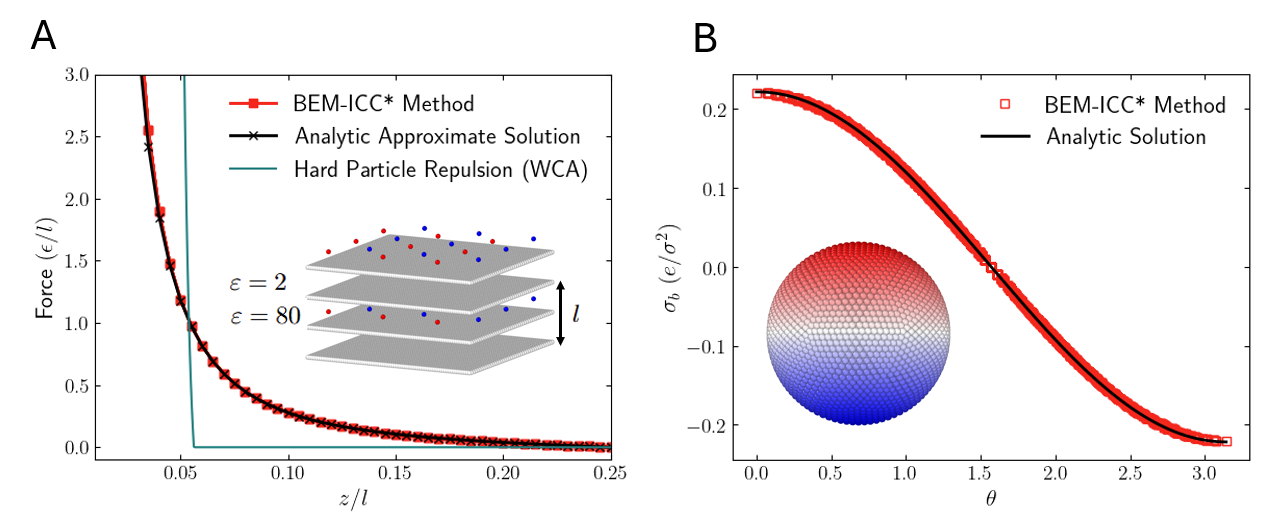}
    \caption{Validation of the new DIELECTRIC package in static cases. (A) Comparison of an approximate analytic result with the BEM-ICC* computed results for a particle in a lattice of alternating charges confined between dielectric slabs. $z$ is the distance from the lattice of particles to the nearest interface. The mesh spacing is 0.04$l$, and $s=0.05l$ for the WCA hard-particle repulsion, which is the same ratio of mesh spacing to $s$ for wall-particle interactions in our dynamic simulations. (B) Comparison of the analytic result for a sphere in an applied electric field with the BEM-ICC* result. The surface of the sphere has mesh points at the vertices of an icosphere with 2562 points. $\varepsilon_\text{sphere}=40$ and $\varepsilon_\text{space}=1$, with an applied field of $E=1$ in LJ units.}
    \label{fig:StaticTests}
\end{figure}

We start by demonstrating the package's accuracy in some static cases where the results can be compared to analytic solutions. First, we validated the package's results by comparing it to a new analytic result for a lattice of ions with alternating charges confined between dielectric slabs, with periodicity in all directions (Fig \ref{fig:StaticTests}A). The details of the calculated potential and its derivation are included in Section S4 of this Supporting Information. In the plot of this result, we also showed what the magnitude of the WCA force used to represent hard-particle repulsion in our dynamic simulations would be, given the same ratio of mesh point spacing to WCA lengthscale ($s$) that we used in our dynamic simulations without electric field (for the simulations with electric field, the mesh is even finer). There is no visible discrepancy between the analytic and simulated dielectric repulsive forces until $z=0.03l$, at which point the hard-particle repulsion is far larger than the dielectric repulsion, showing that these discrepancies should not affect the simulations.

We validated the slab correction by comparing the results obtained with the slab correction (\verb|kspace_modify slab x|) with the results of simulations where an appropriate amount of empty space is added to the top of the simulation box to match the slab correction for a few values of \verb|x| and we found agreement between the simulations.

Finally, to validate the newly added capability of the package to work with externally applied electric fields, we used the case of a dielectric sphere embedded in a medium with different permittivity under an applied electric field. There is a well-known analytical expression for the surface charge density of a sphere in an applied field\cite{griffiths_introduction_1999}:
\[
\sigma_b(\theta)=\frac{3\varepsilon_0(\varepsilon_\text{sphere}-\varepsilon_\text{space})}{\varepsilon_\text{sphere}+2\varepsilon_\text{space}} E_\text{applied}\cos\theta
\]
which we compared with our data (Fig \ref{fig:StaticTests}B). The agreement is very good.
\subsection{Validation for Dynamic Cases}
\begin{figure}[H]
    \centering
    \includegraphics[width=1\linewidth]{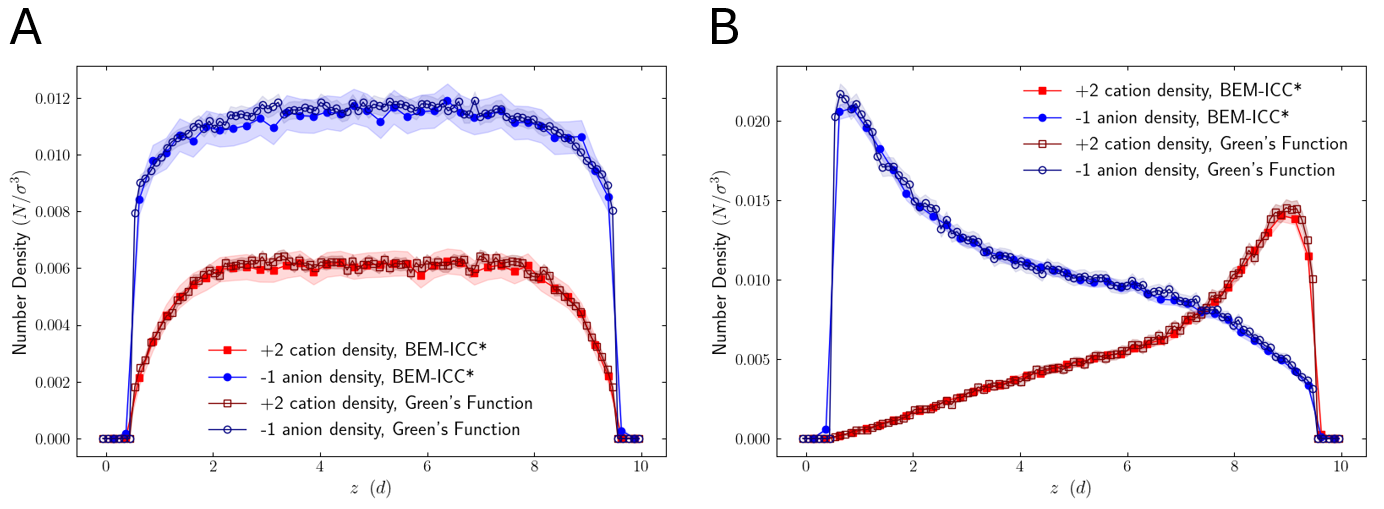}
    \caption{Comparison of ion concentrations from the BEM-ICC* and Green's function based methods for a 2:1 salt solution between planar dielectric interfaces. $z$ is the distance from the lower boundary of the slit. The average concentrations of cations and anions were 0.0056 and 0.011, respectively, in the free volume of the slit (not including the excluded volume near the walls). Shaded regions show 95\% confidence intervals. (A) shows the results with no applied electric field (4500 independent samples), and (B) shows the results with an applied field that corresponds to a field of 24 $k_BT/(e\sigma)$ applied an infinite distance away (7500 independent samples).}
    \label{fig:FlatDynamicTests}
\end{figure}

To further validate the performance of the new version of the DIELECTRIC package, we tested it on a system of monovalent anions and divalent cations confined between two planar polarizable interfaces. The results were compared to results obtained from simulations with a Green's function-based method \cite{pogharian_electric_2024}.  Our system had 20 divalent cations and 40 monovalent anions confined in a slit with a width of 10$\sigma$ in z and infinite extent in the x and y directions. The simulation box had dimensions $L_x=20\sigma$, $L_y=20\sigma$, $L_z=100\sigma$, with periodic boundary conditions in all directions. The dielectric constant inside the slit was $\varepsilon_B=80$, representing water, and was $\varepsilon_A=2$ in the walls. Water was modeled implicitly through NVT Langevin dynamics. We simulated this system both without an applied electric field (Fig \ref{fig:FlatDynamicTests}A) and with an applied field (Fig \ref{fig:FlatDynamicTests}B). In the DIELECTRIC package, the electric field specified in \verb|fix efield/dielectric| is the average electric field in the box. In this system, we chose the electric field to match a system without $z$ periodicity, where the electric field at infinity was $24 k_BT/(e\sigma)$, meaning that ions in the slit felt an electric field of $0.6k_BT/(e\sigma)$.

\begin{figure}[H]
    \centering
    \includegraphics[width=1\linewidth]{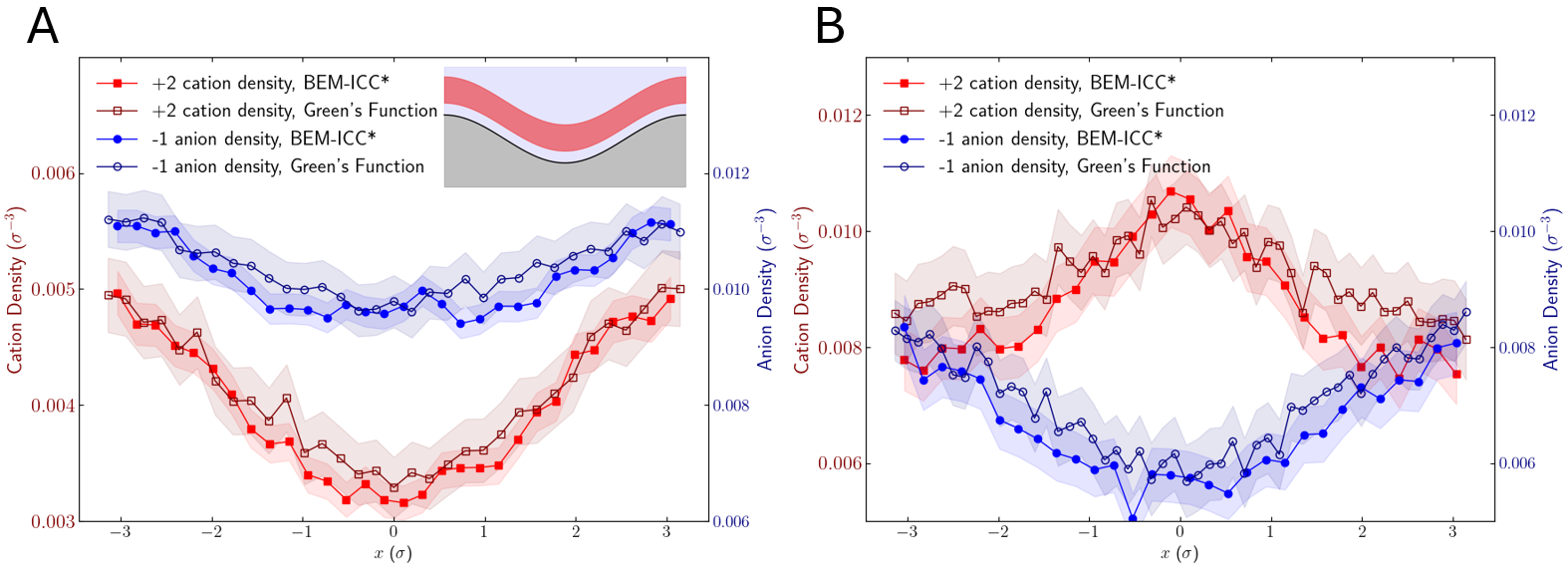}
    \caption{Comparison of ion concentrations near the sinusoidal wall of the slit from the BEM-ICC* and Green's function-based methods for a 2:1 salt solution between planar dielectric. The bulk concentrations of cations and anions were 0.0058 and 0.0116 respectively, in the free volume of the slit. The red shaded region of the inset shows the region where the concentrations are measured: between $0.5\sigma$ and 1.6$\sigma$ from the sinusoidal wall. Shaded regions show 95\% confidence intervals. (A) shows the results with no applied electric field, and (B) shows the results with an average electric field of $E_z=-3.2k_BT/(e\sigma)$. Note that in (A) the cation and anion concentration are plotted on different scales, while in (B) they are plotted on the same scale.}
    \label{fig:SinDynamicTests}
\end{figure}

We also compared the methods on a similar system with a flat wall at the top of the slit and one sinusoidal wall with height $h(x)=0.5\cos(x)$ at the bottom. The interaction of a charge with a dielectric interface depends on the curvature of the interface, so the concentration of ions near the sinusoidal interface varies along its length due to its changing curvature \cite{fj_pimples_2021, wu_asymmetric_2018}. In this simulation, the slit was $10\sigma$ wide on average and the box had dimensions $L_x=6\pi\sigma=18.85\sigma$, $L_y=18.32\sigma$, $L_z=100\sigma$ with periodic boundary conditions in all directions. $L_x$ is three periods of the sinusoid that give the height of the surface, and $L_y$ was chosen so that the hexagonal meshes of adjacent boxes fit together correctly. We used both methods to determine the concentration of ions near the sinusoidal interface, with "near" meaning less than one Debye length above the interface. The Debye length is $l_d=1.6\sigma$, and ions cannot get closer than $0.5\sigma$ to the wall, so this corresponds to the region between $0.5\sigma$ and $1.6\sigma$ above the wall. The results are shown in Fig \ref{fig:SinDynamicTests}A (without applied electric field) and Fig \ref{fig:SinDynamicTests}B (with applied field). Since the sinusoidal wall makes it impossible to exactly compute what field should be used to match a system with a particular electric field at infinity, we simply used an average field of $-3.2 k_BT/(e\sigma)$ over the box, corresponding to a voltage difference of $\phi(z=100)-\phi(z=0)=320k_BT/e$ over the height of the box. The downward field pushes cations toward the sinusoidal interface and anions toward the flat interface.

In both systems, the BEM-ICC* simulation used the WCA potential to model hard-particle repulsion with the walls, as described above, but the Green's function method simulation used reflective walls, where ions are reflected if they come within 0.5$\sigma$ of the wall. For the BEM-ICC* simulations with flat walls, we used a cutoff distance of $10\sigma$ for the real space part of the Coulomb interaction calculations, and a PPPM grid with 8, 8, and 50 points in the x, y, and z directions, respectively. For the system with a sinusoidal wall, we used a cutoff distance of $5.5\sigma$, and a PPPM grid with 16, 16, and 100 points in the x, y, and z directions, respectively.

The agreement with the reference results is excellent for all the tests, except for small differences in the concentration of ions, particularly cations, near the sinusoidal wall in the simulation with an applied electric field (Fig \ref{fig:SinDynamicTests}B). One reason for this may be that the cations are packed together relatively closely near the wall, and the BEM-ICC* method uses a wall made up of a mesh of beads with  WCA repulsive interaction with the ions, while the Green's function method uses a reflective boundary for the walls. Additionally, the methods use different approximations. The BEM-ICC* method has discrete patches of charge, and the Green's function method computes dielectric self interaction terms to first order in the height of the walls.

\section{Section S4. Derivation of Force on a Lattice of Charges}

Consider a set of point charges in a dielectric system composed of
a set of periodic slabs.

The system is periodic in the $x-$direction, with period $L$. In
each period, there are two slab subregions of permittivities $\varepsilon_{A}$
and $\varepsilon_{B}$. Region $A$, where charges are present, spans
the interval from $x=-\alpha L/2$ to $x=\alpha L/2$. Region $B$
spans the rest of space ($|x|\geq\alpha L/2$, in the fundamental
region $|x|\leq L/2$).

\begin{figure}
    \centering
    \includegraphics[width=0.5\linewidth]{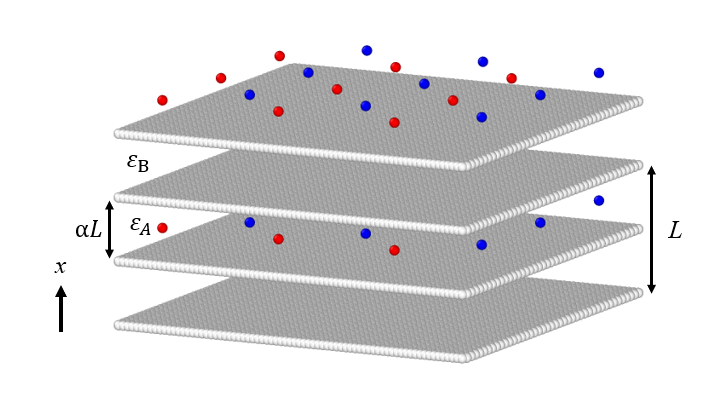}
    \caption{The system considered in Section S4, with a lattice of charges. The blue beads represent negative charges and the red beads positive charges.}
    \label{fig:slab_lattice}
\end{figure}

In each region A, there is a set of charges in a lattice arrangement.
All charges have the same $x-$coordinate so that they exist in a plane
parallel to the dielectric contrast interfaces. A positive charge
can be taken to be at position $\mathbf{a}=(a,0,0)$. Other lattice
charges are located at positions generically labeled $\mathbf{a}_{p}=(a,y_{p},z_{p})$.
The lattice is assumed to be electroneutral (summing charges in its
unit cell), but the limit can be taken to large spacing so that it is
possible to recover results (up to the careful handling of some singularities)
for a single positive charge in each of the $A$ regions. The system (with a lattice of charges) is shown in Fig \ref{fig:slab_lattice}.

To obtain the potentials and fields due to these charges, we need to
solve the Poisson equation

\[
\nabla_{x}\varepsilon_{0}\varepsilon\cdot\nabla_{x}G(\mathbf{x},\mathbf{a})=-\sum_{replicas}\sum_{p\in lattice}q_{p}\delta(\mathbf{x}-\mathbf{a}_{p}).
\]
Below we omit the $\varepsilon_{0}$ factor as it can be incorporated into the relative permittivity of each region. 

The charge distribution can be expressed as 

\[
\rho=\sum_{n\in replicas}\sum q_p\delta(x-a+nL)\delta(y-y_{p})\delta(z-z_{p})
\]

\[
\rho=\sum_{n\in replicas}\delta(x-a+nL)\sum_{p}q_{p}\int\frac{d^{2}k}{(2\pi)^{2}}e^{ik_{y}(y-y_{p})}e^{ik_{z}(z-z_{p})}
\]

\[
\rho=\sum_{n\in replicas}\delta(x-a+nL)\int\frac{d^{2}k}{(2\pi)^{2}}e^{i\mathbf{k}\cdot\mathbf{r}}\sum_{p}q_{p}e^{-i\mathbf{k}\cdot\mathbf{r_{p}}}
\]

\[
\rho=\sum_{n\in replicas}\delta(x-a+nL)\int\frac{d^{2}k}{(2\pi)^{2}}e^{i\mathbf{k}\cdot\mathbf{r}}u(\mathbf{k})
\]
Here we have introduced the two-dimensional vectors $\mathbf{r}=(y,z)$,
and $\mathbf{k}=(k_{y},k_{z})$. The sum of the exponential factor
over the position of the lattice points in a single transverse plane,
$u(\mathbf{k})$, is the reciprocal space density of the lattice, and
is formally a sum of delta functions in $k$-space 

\[
u(\mathbf{k})=\sum_{\mathbf{k}_{p'}}u_{p'}\delta(\mathbf{k}-\mathbf{k}_{p'}).
\]
However, it is possible to consider the limit of a single charge at
the origin of the transverse lattice. In that case

\[
u(\mathbf{k})=1.
\]

To match the form of the Green's function to that of the charge term,
we seek a solution of the form 

\[
G(\mathbf{x},\mathbf{a})=\int d^{2}k\,\frac{1}{(2\pi)^{2}}g(x,\mathbf{k};a)e^{ik_{y}y}e^{ik_{z}z}.
\]
Plugging this expression into the Poisson equation, we see that the function $g$ satisfies

\[
\partial_{x}\varepsilon_{0}\varepsilon\partial_{x}g-\varepsilon_{0}\varepsilon k^{2}g=-u(\mathbf{k})\delta(x-a)
\]
in the fundamental region, with periodic boundary conditions. In each
of the homogeneous regions with respect to the horizontal coordinate,
where there is no boundary or charges, solutions are combinations
of exponentials $e^{\pm kx}$ or of hyperbolic functions. 

The delta function can be matched by using a contribution $g_{\delta}$,
within the $A$ region, to the function $g$, of the form 

\[
g_{\delta}=\frac{u(\mathbf{k})}{2\varepsilon_{A}k}\begin{cases}
\exp(k(x-a)) & -\alpha L/2\leq x<a\\
\exp(-k(x-a)) & \alpha L/2\geq x>a
\end{cases}
\]
The total solution can be written as

\[
g=g_{\delta}+h_{A}+h_{B}
\]
where $h_{A,B}$ are solutions of the homogeneous equation in the
$A$ and $B$ regions and are zero elsewhere. We can take

\[
h_{A}=A'\cosh kx+B'\sinh kx,\;\;\;\;|x|\leq\alpha L/2
\]

\[
h_{B}=C'\cosh(k(x-L/2))+D'\sinh(k(x-L/2)),\;\;\;\;\alpha L/2<x<L-\alpha L/2,
\]
 and its replicates in all other locations.

The equations that result from boundary conditions all have inhomogeneous
terms that are proportional to $u(\boldsymbol{k})$, due to a factor of $u(\boldsymbol{k})$ in $g_\delta$. We factorize this value and
obtain the conditions for 

\[
A=u(\mathbf{k})A',\;\;\;\;B=u(\mathbf{k})B',\;\;\;\;C=u(\mathbf{k})C',\;\;\;\;D=u(\mathbf{k})D'.
\]

At the interfaces, we have the following conditions. Continuity at
$x=-\alpha L/2$; $h_{A}$ and $g_{\delta}$ are evaluated there and
we match the limiting value of $h_{B}$ at $L-\alpha L/2$

\begin{eqnarray*}
A\cosh k\alpha L/2-B\sinh k\alpha L/2+\frac{1}{2\varepsilon_{A}k}\exp[-k(a+\alpha L/2)]\\
=C\cosh(k(1-\alpha)L/2)+D\sinh (k(1-\alpha) L/2)
\end{eqnarray*}
Continuity at $x=\alpha L/2$ gives

\begin{eqnarray*}
A\cosh k\alpha L/2+B\sinh k\alpha L/2+\frac{1}{2\varepsilon_{A}k}\exp[-k(\alpha L/2-a)]\\
=C\cosh(k(1-\alpha) L/2)-D\sinh(k(1-\alpha)L/2).
\end{eqnarray*}
Matching the horizontal component of the displacement vector $\varepsilon\partial_{x}G$
at $x=-\alpha L/2$ gives

\begin{eqnarray*}
-\varepsilon_{A}kA\sinh k\alpha L/2+\varepsilon_{A}kB\cosh k\alpha L/2+\frac{1}{2}\exp[-k(a+\alpha L/2)]\\
=\varepsilon_{B}kC\sinh (k(1-\alpha) L/2)+\varepsilon_{B}Dk\cosh (k(1-\alpha) L/2)
\end{eqnarray*}

Matching of the displacement vector at $x=\alpha L/2$ gives

\begin{eqnarray*}
\varepsilon_{A}kA\sinh k\alpha L/2+\varepsilon_{A}kB\cosh k\alpha L/2-\frac{1}{2}k\exp[-k(\alpha L/2-a)]\\
=-\varepsilon_{B}Ck\sinh(k(1-\alpha) L/2)+\varepsilon_{B}Dk\cosh(k(1-\alpha) L/2).
\end{eqnarray*}

We can write this system as

\[
Mw=v
\]

with 

\[
M=\left(\begin{array}{cccc}
\cosh k\alpha L/2 & -\sinh k\alpha L/2 & -\cosh (k(1-\alpha)L/2) & -\sinh (k(1-\alpha)L/2)\\
\cosh k\alpha L/2 & \sinh k\alpha L/2 & -\cosh (k(1-\alpha)L/2) & \sinh (k(1-\alpha)L/2)\\
-\varepsilon_{A}k\sinh k\alpha L/2 & \varepsilon_{A}k\cosh k\alpha L/2 & -\varepsilon_{B}k\sinh (k(1-\alpha)L/2) & -\varepsilon_{B}k\cosh (k(1-\alpha)L/2)\\
\varepsilon_{A}k\sinh k\alpha L/2 & \varepsilon_{A}k\cosh k\alpha L/2 & \varepsilon_{B}k\sinh (k(1-\alpha)L/2) & -\varepsilon_{B}k\cosh(k(1-\alpha)L/2)
\end{array}\right),
\]

\[
w=\left(\begin{array}{c}
A\\
B\\
C\\
D
\end{array}\right),v=\left(\begin{array}{c}
-\frac{1}{2\varepsilon_{A}k}\exp[-k(a+\alpha L/2)]\\
-\frac{1}{2\varepsilon_{A}k}\exp[k(a-\alpha L/2)]\\
-\frac{1}{2}\exp[-k(a+\alpha L/2)]\\
\frac{1}{2}\exp[k(a-\alpha L/2)]
\end{array}\right)
\]

The solution to the system is schematically

\[
w=M^{-1}v.
\]
Explicit expressions can be obtained by symbolic manipulation. The result
is that the coefficients are expressed as ratios of sums of products
of exponential and hyperbolic functions of the magnitude of $k.$
The result is also linear in $u(\mathbf{k})$ for a given $\mathbf{k}$
value. 

This completes the determination of the Green's function for the set
of charges considered. It can be checked that the solutions obtained
for the coefficients are rapidly decaying functions of the reciprocal
vector.

The function obtained, $G(\mathbf{x},\mathbf{a})$, is the potential
at any point in space. It can be used to obtain the field and the
force on a test charge. However, the force on a test charge in a
periodic cell has to be carefully defined since including the particle
in the system introduces its images as well. One approach to avoid
this issue is to consider the force on the same particle located at
$\mathbf{a}.$ The contributions of its replicas have already been considered, but it will be necessary to subtract any self interaction terms. 

To be able to subtract self-interaction terms, we match the inhomogeneous
part of the solution to the Green's function of a single particle. Following the steps above for an infinite length $L$ and a single particle in a medium of type A, we have the Green's function of the single particle as

\[
G^{0}(\mathbf{x},\mathbf{a})=\frac{1}{4\pi\varepsilon_{A}}\frac{1}{|\mathbf{x}-\mathbf{a}|}=\int d^{2}k\,\frac{1}{(2\pi)^{2}}g^{0}(x,\mathbf{k};a)e^{i\mathbf{k}\cdot\mathbf{r}}
\]

with 

\[
g^{0}=g_{\delta}=\frac{1}{2\varepsilon_{A}k}\begin{cases}
\exp k(x-a) & x\leq a\\
\exp(-k(x-a)) & x>a
\end{cases}
\]
Now, the potential at the position $\mathbf{a}$ is the formal evaluation
of the Green's function at that location

\[
G(\mathbf{a},\mathbf{a})=\int d^{2}k\,\frac{1}{(2\pi)^{2}}g(a,\mathbf{k};a).
\]

\[
G(\mathbf{a},\mathbf{a})=\int d^{2}k\,\frac{u(\mathbf{k})}{2k\varepsilon_{A}(2\pi)^{2}}+\int d^{2}k\,\frac{u(\mathbf{k})}{(2\pi)^{2}}[A(\mathbf{k})\cosh(ka)+B(\mathbf{k})\sinh(ka)].
\]
Subtracting the self-contribution, we obtain the effective potential
$G'=G-G^{0}(\mathbf{a},\mathbf{a})$.

. 
\[
G'(\mathbf{a},\mathbf{a})=\int d^{2}k\,\frac{[u(\mathbf{k})-1]}{2k\varepsilon_{A}(2\pi)^{2}}+\int d^{2}k\,\,\frac{u(\mathbf{k})}{(2\pi)^{2}}[A(\mathbf{k})\cosh(ka)+B(\mathbf{k})\sinh(ka)].
\]

In the case of the single charge in region $A$, the first term is
zero, but it can be shown that the second term has a logarithmic singularity.
Nevertheless, the expression can be used to obtain the horizontal
electric field, which is regular. On the other hand, when the set
of charges is an electroneutral lattice, the first integral can be
given meaning by a limiting procedure, while the second one is finite. 

The $x$-component of the electric field created by the charges
is 

\[
E_{x}(\mathbf{x},\mathbf{a})=-\partial_{x}G(\mathbf{x},\mathbf{a})=-\int d^{2}k\,\frac{1}{(2\pi)^{2}}\partial_{x}g(x,\mathbf{k};a)e^{ik_{y}y}e^{ik_{z}z}.
\]

When evaluating at $\mathbf{x}=\mathbf{a}$, it can be checked that it
is possible to assign the value of 0 to the derivative of $g_{\delta}$
and to the term arising from $G^{0}.$ Therefore, the field at this
point is

\[
E_{x}(\mathbf{a},\mathbf{a})=-\int d^{2}k\,\,\frac{u(\mathbf{k})}{(2\pi)^{2}}[A(\mathbf{k})k\sinh(ka)+B(\mathbf{k})k\cosh(ka)].
\]
The integral is convergent in the case of the single particle as well
as for an electroneutral lattice. 

For the single particle we obtain, 

\[
E_{x}(\mathbf{a},\mathbf{a})=-\int_{0}^{\infty}\frac{1}{2\pi}kdk\,[A(k)k\sinh(ka)+B(k)k\cosh(ka)].
\]

For the case of a lattice we obtain, using the fact that $u(\mathbf{k})=\sum_{\mathbf{k}_{rec}}u_{p'}\delta(\mathbf{k}-\mathbf{k}_{p'})$, 

\[
E_{x}(\mathbf{a},\mathbf{a})=-\sum_{p}\frac{1}{(2\pi)^{2}}u(\mathbf{k}_{p})[A(\mathbf{k}_{p})k_{p}\sinh(k_{p}a)+B(\mathbf{k}_{p})k_{p}\cosh(k_{p}a)]
\]

Consider the case of a lattice with positive charges at $y=dn$, $z=dm$,
with $n,m$ integers, and with negative charges at $y=d(n+1/2)$,
$z=d(m+1/2)$ for $n,m$ integers. Use as unit cell the square $|y|\leq1/2$,
$|z|\leq1/2$. There is one point charge in the center and four $1/4$
negative charges at the corners. A Fourier series of the density can
be written as 

\[
\rho(\mathbf{r})=\sum_{p}s_{_{p}}e^{i\mathbf{k}_{p}\cdot\mathbf{r}}
\]
 where the sum is over the reciprocal vectors $\mathbf{k}_{p}=(2\pi n/d,2\pi m/d).$ The
Fourier coefficients are:

\[
s_{_{p}}=\frac{1}{d^{2}}\int_{unit\,cell}e^{-i\mathbf{k}_{p}\cdot\mathbf{r}}\rho(\mathbf{r})d\mathbf{r}
\]

They evaluate to 

\[
s_{p}=\frac{1}{d^{2}}(1-\frac{1}{4}e^{-i\pi n(n+m)}-\frac{1}{4}e^{-i\pi n(n-m)}-\frac{1}{4}e^{-i\pi n(-n+m)}-\frac{1}{4}e^{-i\pi n(-n-m)})
\]

\[
s_{p}=\frac{1}{d^{2}}(1-cos(\pi n)\cos(\pi m))=\frac{1}{d^{2}}(1-(-1)^{n+m}).
\]

Therefore, the Fourier transform of the density can be written as

\[
u(\mathbf{k})=(2\pi)^{2}\sum_{\mathbf{k}_{rec}}s_{p'}\delta(\mathbf{k}-\mathbf{k}_{p'}).
\]

The field is 

\[
E_{x}(\mathbf{a},\mathbf{a})=-\sum s_{p}[A(k_{p})k_{p}\sinh(k_{p}a)+B(k_{p})k_{p}\cosh(k_{p}a)]
\]

\[
E_{x}(\mathbf{a},\mathbf{a})=-\frac{1}{d^{2}}\sum_{n,m}(1-(-1)^{n+m)})[A(k_{p})k_{p}\sinh(k_{p}a)+B(k_{p})k_{p}\cosh(k_{p}a)]
\]

with $k_{p}=2\pi\sqrt{(n^{2}+m^{2})}/d.$

\section{Section S5. Mesh Validation}

\begin{figure}[H]
    \centering
    \includegraphics[width=0.9\linewidth]{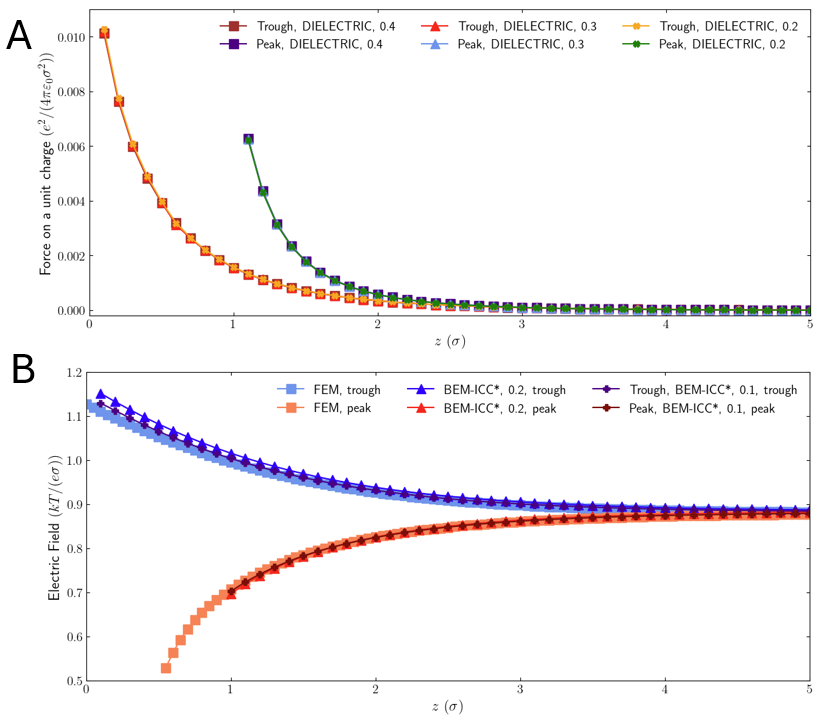}
    \caption{Mesh Validation. In both figures, $z$ is the distance from the position of a test particle to the midpoint of a sinusoidal wall with height given by $z=h(x)=0.5\cos(x)$. Due to the repulsive WCA interactions used to simulate steric repulsion, the ions cannot approach closer than $\approx0.55\sigma$ to the local surface height before the WCA repulsion dwarfs the dielectric self interaction. Trough means lowest point of the cosine ($x=\pi$) and peak means the highest point ($x=0$). (A) Validation for the mesh without an electric field. Mesh spacings of 0.2$\sigma$, 0.3$\sigma$, and 0.4$\sigma$ were tested without an electric field, with permittivity 80 in the medium containing the charge and permittivity 2 in the walls. The difference in force between a spacing of $0.2\sigma$ and $0.4\sigma$ was $1.5\%$, and we deemed the $0.4\sigma$ mesh sufficiently accurate. (B) Validation for the mesh with an electric field. We compared the electric field on a charged particle with electric field results from an FEM simulation using the SfePy Python package\cite{Cimrman_Lukes_Rohan_2019}. Mesh spacings of 0.2$\sigma$, 0.3$\sigma$, and 0.4$\sigma$ were tested, with permittivity 80 in the medium containing the charge and permittivity 8 in the walls. For a mesh spacing of $0.1\sigma$, the discrepancy with the FEM simulation was $1.4\%$, which we deemed sufficiently accurate.}
    \label{fig:MeshConvergence}
\end{figure}

To compute the induced charge on dielectric interfaces in our MD simulations, we discretize the interfaces into a hexagonal mesh of points. For the simulations with no applied electric field, we use a dielectric permittivity of 80 for the solvent and 2 for the wall, with a mesh spacing of 0.4$\sigma$. For the simulations with an applied electric field, we found that a ratio of 40 between the dielectric constants required a very fine mesh that was untenably computationally intensive, so we used a wall with a dielectric permittivity of 8 instead (the permittivity of the solvent was still 80). Validation of our meshes is shown in Fig \ref{fig:MeshConvergence}.

\section{Section S6. Exact Simulation Box Sizes and Concentrations}
Due to the hexagonal meshing and different wavelengths used, in combination with needing to have an integer number of ions per box, the concentration varies slightly between simulations. The exact simulation box sizes and concentrations are given in Table \ref{tab:dimensions}.
\begin{table}[H]
\centering
\begin{tabular}{|l|l|l|l|l|l|} 
\hline
System & $L_x\;(\sigma)$ & $L_y\;(\sigma)$ & $L_z\;(\sigma)$ & Molarity (mM) & $\varepsilon_\text{solvent}-\varepsilon_\text{walls}$\\ 
\hline
\multicolumn{6}{|c|}{Electric Field Response} \\
\hline
$A=0.5,\;k=0.6$ & $10\pi/3$ & 11.09 & 100 & 25.0:50.0 & 80-8\\ 
\hline
$A=0.5,\;k=0.8$ & $5\pi$ & 15.86 & 100 & 25.3:50.5 & 80-8\\ 
\hline
$A=0.5,\;k=1$ & $4\pi$ & 12.22 & 100 & 25.2:50.4 & 80-8\\ 
\hline
$A=0.5,\;k=1.2$ & $10\pi/3$ & 11.05 & 100 & 25.1:50.2 & 80-8\\ 
\hline
$A=2,\;k=1$ & $4\pi$ & 12.17 & 100 & 25.3:50.6 & 80-8\\ 
\hline
\multicolumn{6}{|c|}{Fourier Superposition}\\
\hline
$h(x)=0.25\cos(x)$ & $4\pi$ & 11.74 & 100 & 26.2:52.5 & 80-2\\ 
\hline
$h(x)=0.25\cos(1.5x)$ & $4\pi$ & 12.28 & 100 & 25.1:50.2 & 80-2\\ 
\hline
$h(x)=0.25\cos(x)+0.25\cos(1.5x)$ & $4\pi$ & 12.12 & 100 & 25.4:50.8 & 80-2\\ 
\hline
$h(x)=0.5\cos(x)$ & $4\pi$ & 12.21 & 100 & 25.2:50.5 & 80-2\\ 
\hline
$h(x)=0.5\cos(1.5x)$ & $4\pi$ & 12.28 & 100 & 25.1:50.2 & 80-2\\ 
\hline
$h(x)=0.5\cos(x)+0.5\cos(1.5x)$ & $4\pi$ & 12.06 & 100 & 25.6:51.1 & 80-2\\ 
\hline
\end{tabular}
\caption{Exact Simulation Box Sizes and Concentrations for the MD simulations. The first column identifies the system being described, the next three columns give the dimensions of the system in ion diameters, the fifth column gives the molarity in mM in the form cation molarity:anion molarity, and the final column gives the dielectric permittivities of the solvent and the walls.}
\label{tab:dimensions}
\end{table}
\section*{Section S7. Quantification of Uncertainty for Critical Electric Fields}
Fig 2 in the main text shows the critical electric field to cancel out the modulation in potential due to the curvature of the sinusoidal surface for several wavelengths. However, the electric field is not measured directly from a single simulation; rather, for each simulation, a single, exact value of the electric field is set and the resulting ion distribution is measured. The critical field $E_c$ is reached when the second mode (the first non-constant mode) of the Fourier transform of the cation distribution equals zero. This is the component of the Fourier transform, which we denote $A^{(1)}$, that corresponds to the same wavelength as the wavelength of the surface and that Fourier component is expected to be proportional to the difference between the applied electric field and the critical field strength. By choosing two to three electric fields that produce ion distributions with $A^{(1)}$ close to zero, with at least one on either side of zero, we can estimate the field $E_c$ where $A^{(1)}=0$ using weighted least squares regression, where the weight of each point is $1/\sigma^2_{A^{(1)}}$, with $\sigma_{A^{(1)}}$ the uncertainty in $A^{(1)}$, for a particular value of the Electric field \cite{asuero_fitting_2007}. To find $1/\sigma^2_{A^{(1)}}$, we consider an ion density $\rho(x)$ with estimated standard error $\sigma_\rho(x)$, which is approximated by a binned ion density $\rho_n$, where the density for each bin has uncertainty $\sigma_{\rho_n}$. We estimate the uncertainty in the Fourier transform to be
$$
\sigma_{A_k}^2=\sum_{n=0}^{N-1}\cos^2(2\pi kn/N)\sigma_{\rho_n}^2,
$$
since the symmetry of the troughs means the Fourier transforms should be real. For this calculation, we assumed that the ion concentrations at different x positions are independent. While this assumption is not strictly true - clearly an ion alters the potential around itself and thus the probability of another ion being nearby - the low density of ions means that it is unlikely for two ions to be close enough for this effect to be significant.

The weighted least squares regression gives an approximation for $A^{(1)}$ as a function of applied field $E$: $A^{(k)}(E)=mE+E_0$, with $m$ the slope of the linear approximation and $E_0$ the intercept. The critical field is then $E_c=-E_0/m$. By linearizing $E_c$ in terms of $E_0$ and $m$, the uncertainty in the critical field is estimated to be\cite{taylor_311_1997}:
$$
\sigma_{E_c}=\sqrt{\frac{1}{m^2}\sigma^2_{E_0}+\frac{E_0^2}{m^4}\sigma^2_m-\frac{E_0}{m^3}\text{Cov}(E_0,m)},
$$
where $\sigma_{E_0}$ and $\sigma_m$ are the uncertainties in $E_0$ and $m$ respectively, obtained from the weighted least squares linear regression \cite{asuero_fitting_2007}.

\section*{Section S8. Importance of Dielectric Contrast}

\begin{figure}[H]
    \centering
    \includegraphics[width=1\linewidth]{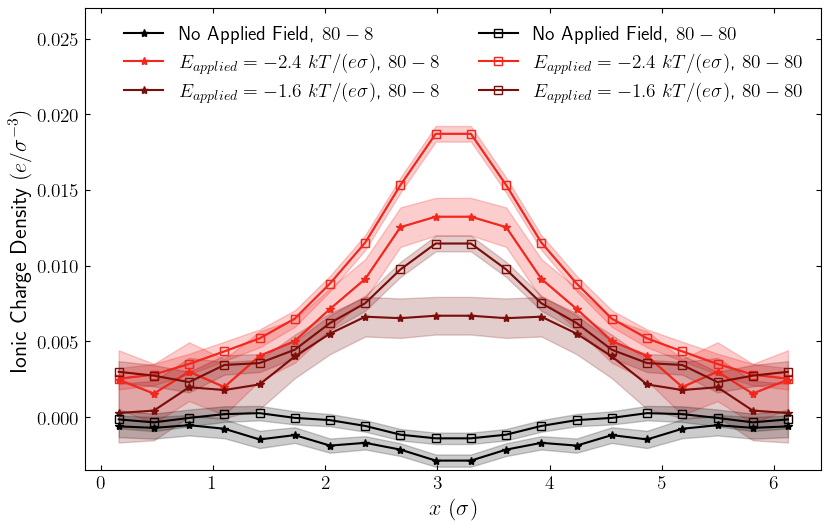}
    \caption{Comparison of the net ionic charge near a sinusoidal surface with an applied electric field, with and without dielectric contrast. This plot shows net ionic charge density near an interface with height $h(x)=2\cos(x)$. The electrolyte is a 2:1 salt with 25 mM divalent cation concentration and 50 mM monovalent anion concentration. The electric field values given here are what the average value of the field would be inside the slit \emph{if there were no screening from the electrolyte}. This allows for a comparison where the only difference between the two systems is the self interaction of ions with the dielectric interface. The shaded regions mark the area within 2 standard errors of the mean charge density at each point.}
    \label{fig:no_contrast}
\end{figure}

In this section, we compare the effect of an electric field on ion distributions near a sinusoidal surface with and without dielectric contrast. Rather than having a solvent with $\varepsilon=80$ and walls with $\varepsilon=8$ or $2$, we use a permittivity of $\varepsilon=80$ for both the walls and the solvent. The net ionic charge distribution near the surface under these conditions, as well as a comparison to the results with dielectric contrast (which are also shown in the main text), is shown in Fig \ref{fig:no_contrast}. The ion distributions are qualitatively similar, but noticeably and quantitatively different, especially near the trough of the sinusoid. The magnitude of the negative charge density without an applied field is also significantly less, which is consistent with the results of Wu et al.\cite{wu_asymmetric_2018}.  Our results show that the dielectric contrast has a significant effect on ion distributions, even with an applied electric field.

\section*{Section S9. Computing modulation in ion density due to multiple Fourier modes}
In the main text, we give the expression $c_\text{sum}(x)=-(N-1)c_{k=0}(x)+\sum_i^N c_{k_i}(x)$ for the predicted near surface ion density due to the presence of multiple Fourier modes in terms of the near-surface ion density due to each mode. This expression is correct if the bulk concentration for the system with each wavelength is the same. However, in our simulations, the length of the simulation box in the y direction varies slightly for different wavelengths, because we are using a hexagonal mesh, and the  y length of the box must be an integer number of interface mesh points so that periodic boundary conditions can be used. The volume of the box varies proportionally with the discrepancies in y length. Due to the relatively small number of ions (8 cations and 16 anions) in the simulation box, these volume differences cannot be compensated for by adding or removing atoms to keep the concentration constant. Since the differences in concentration are less than 4.2\%, we assume that the ion density near the surface is proportional to the bulk concentration, and use the following modified expression for the predicted ion density for a superposition of modes:
$$
c_\text{sum}(x)=\frac{C^\text{bulk}_\text{sum}}{C^\text{bulk}_{k=k_0}}c_{k=0}(x)+\sum_i^N \Big(\frac{C^\text{bulk}_\text{sum}}{C^\text{bulk}_{k=k_i}}c_{k_i}(x)-\frac{C^\text{bulk}_\text{sum}}{C^\text{bulk}_{k=k_0}}c_{k=0}(x)\Big),
$$
where $C^\text{bulk}_\text{sum}$ is the bulk concentration of ions for the simulation with the superposition of modes for which the ion density profile is to be predicted, and $C^\text{bulk}_{k=k_i}$ is the bulk concentration of ions for the simulation with wavelength $2\pi/k_i$.

\section*{Section S10. Importance of Electrolyte Asymmetry}

\begin{figure}[H]
    \centering
    \includegraphics[width=1\linewidth]{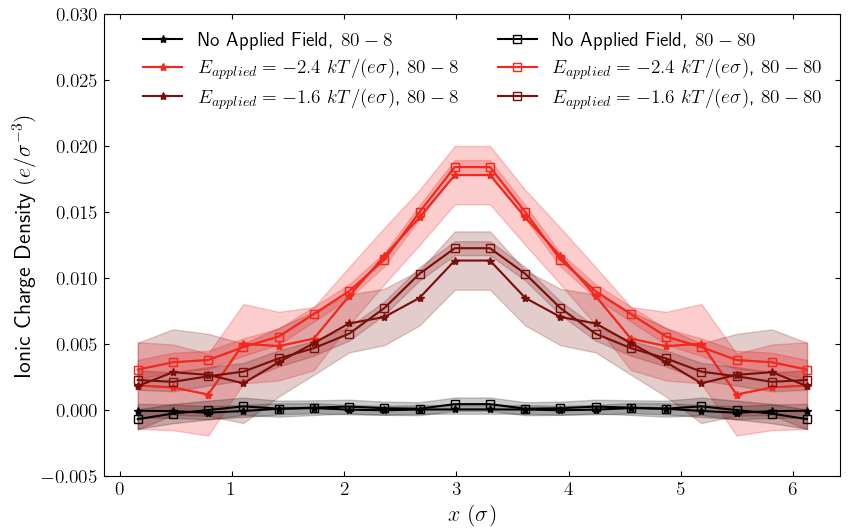}
    \caption{Comparison of the net ionic charge near a sinusoidal surface with an applied electric field, with and without dielectric contrast for a 1:1 electrolyte. This plot shows net ionic charge density near an interface with height $h(x)=2\cos(x)$. The electrolyte is a 1:1 salt at a concentration of 75 mM, so chosen because this solution has the same Debye length as for the 2:1 25:50 mM salt we used in the rest of the paper. The electric field values given here are what the average value of the field would be inside the slit \emph{if there were no screening from the electrolyte}. The shaded regions mark the area within 2 standard errors of the mean charge density at each point.}
    \label{fig:1-1}
\end{figure}

In this section, we show the ionic charge density near a surface in a 1:1 electrolyte, demonstrating that a 2:1 electrolyte produces more complex and interesting behavior. The net ionic charge density for a 1:1 electrolyte near a sinusoidal wall, both with and without dielectric contrast, is shown in Fig \ref{fig:1-1}. Without an applied electric field, the charge density is zero across the sinusoid, since the cation and anion concentrations near the surface are the same. Even with an applied field, the cases with and without dielectric contrast are not significantly different. We would expect the ionic charge density to be slightly lower in the trough in the case with dielectric contrast, because the cations that are pushed into the trough by the electric field will be repelled by the dielectric contrast, reducing the cation density, whereas the anion density will be affected less by the dielectric contrast since it is already low in the trough due to the effect of the field. This difference is not statistically distinguishable in the MD simulation data and we did not run our simulations for long enough to resolve this difference it because is only tangentially related to the rest of the paper.

\bibliography{bibliography}

\end{document}